  \documentclass[aps,preprint,superscriptaddress,showpacs]{revtex4-1}
\usepackage{multirow}
\usepackage{graphicx}

\begin{document}
\bibliographystyle{apsrev}
\title{ Partial-wave analysis of $\pi^- p$ $\rightarrow$  $\eta n$ and $\pi^-p\rightarrow K^0\Lambda$ reactions}
\author{M. Shrestha}
\author{D.~M.~Manley}
\affiliation{Department of Physics, Kent State University, Kent, OH 44242-0001}

\begin{abstract}
We investigate the hadronic reactions $\pi N\rightarrow \eta N$ and $ \pi N \rightarrow K\Lambda$ via single-energy partial-wave analyses in the c.m.\ energy range 1080 to 2100 MeV. Our results for the $K\Lambda$ channel are consistent with prior works; however, for the $\eta N$ channel our results differ significantly from previous energy-dependent partial-wave analyses that violate the $S$-matrix unitarity. We present the first (new) results of $\eta N$ and $K\Lambda$ partial-wave amplitudes constrained by a unitary energy-dependent model.  We obtain excellent predictions of integrated cross sections for the two reactions from a global energy-dependent solution. Our results imply that the region just above $S_{11}(1535)$ has a major contribution from $P_{11} (1710)$ for $\pi^-p\rightarrow \eta n$, whereas the large peak near 1700 MeV in $\pi^-p\rightarrow K^0\Lambda$ is dominated by contributions from both $S_{11} (1650)$ and $P_{11} (1710)$. 


\end{abstract}
\maketitle
\section{Introduction and Motivation} 

         The importance of the hadronic reactions $\pi N \rightarrow \eta N$ and $\pi N\rightarrow K\Lambda$ cannot be overstated. The huge amount of high-quality data on the electromagnetic processes $\gamma N \rightarrow \eta N$ and $\gamma N \rightarrow K\Lambda$ from various facilities (ELSA, GRAAL, JLAB, LEPS, MAMI), when analyzed and interpreted by phenomenologists, will certainly lead to a clearer picture of the baryon resonance spectrum.
 The validity of resonance parameters thus extracted will not be substantiated without similar results from studies of the corresponding hadronic reactions. The study of $\pi N \rightarrow \eta N$ and $\pi N\rightarrow K\Lambda$ complements the study of eta and kaon photoproduction. 
 
     Most previous partial-wave analyses (PWAs) of $\pi^-p\rightarrow \eta n$  \cite{feltesse75, baker79} and $\pi^-p\rightarrow K^0\Lambda$ \cite{knasel75, baker78, saxon80, bell83} were based on the assumption that  partial-wave amplitudes could be represented by a simple sum of resonant and  background terms. Such an assumption violates  unitarity of the partial-wave $S$-matrix. In this work, we report on our investigation of the reactions $\pi^-p\rightarrow \eta n$ and $\pi^-p\rightarrow K^0\Lambda$ via single-energy analyses.  All available differential cross section, polarization, polarized cross section, and spin-rotation data within the energy limits of this analysis were fitted. In order to ensure that our amplitudes had a relatively smooth variation with energy, we introduced several constraints that will be described in detail below.

   \section{Formalism and Fitting Procedures}
        Here, we summarize the formalism for the single-energy partial-wave analyses. The data were analyzed in small energy bins.  Within each energy bin, each amplitude was approximated as a complex constant.
        The differential cross section $\rm d\sigma /\rm d\Omega$ and polarization $P$ are given by
               \begin{equation}
               \frac{{\rm d}\sigma}{{\rm d}\Omega} = {\lambdabar}^2(|f|^2+|g|^2)~,
                \end{equation}
                \begin{equation}
                P\frac{{\rm d}\sigma}{{\rm d}\Omega} =2{\lambdabar}^2\rm Im(fg^\ast)~,
                \end{equation}
        where $\lambdabar = {\hbar}/{k}$,
         with $k$ the magnitude of c.m.\ momentum of the incoming particle.
         In addition, the spin-rotation parameter is defined by
         \begin{equation}
         \beta = \arg\left (\frac{f-ig}{f+ig}\right ),
         \end{equation}
         from which it follows that
         \begin{equation}
         \beta = \tan^{-1}\left (\frac{-2{\rm Re}(f^*g)}{|f|^2 - |g|^2}\right ).
         \end{equation}
         Here, $f = f(W,\theta)$ and $g = g(W,\theta)$ are the usual spin-non-flip and spin-flip amplitudes at c.m.\ energy $W $ and meson c.m.\ scattering angle $\theta$. In terms of partial waves, $f$ and $g$ can be expanded as
       
       \begin{equation}
        f(W,\theta) = \sum_{l=0}^{\infty} [(l+1)T_{l+} + lT_{l-}]P_l(\cos\theta)~,
        \end{equation}
        \begin{equation}
        g(W,\theta) = \sum_{l=1}^{\infty} [T_{l+} - T_{l-}]P_l^1(\cos\theta)~,
        \end{equation}
        where $l$ is the initial orbital angular momentum, $P_l(\cos\theta)$ is a Legendre polynomial and $P_l^1(\cos\theta) = \sin\theta \cdot {\rm d} P_l(\cos\theta)/{\rm d}(\cos\theta$). The total angular momentum for the amplitude $T_{l+}$ is $J=l+\frac12$, while that for the amplitude $T_{l-}$ is $J=l-\frac12$.
              For the initial $\pi N$ system, we have $I = 1/2$ or $I =3/2$ so that the amplitudes $T_{l\pm}$ can be expanded in terms of  isospin amplitudes as 
         \begin{equation}
               T_{l\pm} = C_{\frac12}T^{\frac12}_{l\pm} + C_{\frac32}T^{\frac32}_{l\pm}~,
       \end{equation}
       \newline
       where $T^I_{l\pm}$ are partial-wave amplitudes with isospin $I$  and total angular momentum $J = l\pm\frac12$ with $C_I$ the appropriate isospin Clebsch-Gordon coefficients for a given reaction.
       For $\pi^-p\rightarrow \eta n$ and $\pi^-p\rightarrow K^0\Lambda$, we have $C_{\frac12} = -\sqrt{\frac23}$ and $C_{\frac 32}=0$.
\newline


        
          Single-energy fits were performed separately for the  two reactions  $\pi^- p \rightarrow \eta n$   and $\pi^- p\rightarrow K^0 \Lambda$.
 In each case the available data were analyzed in  c.m.\ energy bins of width 30 MeV. This choice of bin width was appropriate because the data for smaller widths had unacceptably low  statistics and for larger  widths, some amplitudes varied too much to approximate them as constants over the energy spread of the bin. 
  
  Tables I and II summarize  the available quantity and types of data in each energy bin for the two inelastic reactions. Spin-rotation-parameter data were available only for $\pi^-p \rightarrow K^0\Lambda$ and no data at all were available for $\pi^-p\rightarrow \eta n$ in the bins centered at $W$ = 1740, 1800, 1950, and 2040 MeV.  

\begin{table}[htpb]
\caption{Statistics for single-energy fits for $\pi^- p\rightarrow\eta$ $n$.}
\begin{center}
\begin{tabular}{|c|c|c|c|}
\hline
$W$(MeV) &${\rm d\sigma}/{\rm d\Omega}$ & $P$  & References\\ \hline
$1530\pm15$ &89  &--  & \cite{prakhov05, debeham75, deinet69, feltesse75}  \\
$1560\pm15$ &47  &--  & \cite{richards70, debeham75, brown79}  \\
$1590\pm15$ &43  &--  &  \cite{richards70, debeham75, deinet69}  \\    
$1620\pm15$ &28   &--  &  \cite{debeham75, deinet69}   \\
$1650\pm15$ &15   &--  &  \cite{debeham75, brown79}    \\
$1680\pm15$ &45   &--  &  \cite{debeham75, brown79}    \\
$1710\pm15$ &18   &--  &  \cite{brown79}    \\
$1740\pm15$ &--   &--  &      \\
$1770\pm15$ &19   &5 & \cite{brown79, baker79}     \\
$1800\pm15$ &--   &-- &       \\
$1830\pm15$ &19   &5 & \cite{brown79, baker79}    \\
$1860\pm15$ &20   &7 & \cite{brown79, baker79}    \\
$1890\pm15$ &20   &6 &  \cite{brown79, baker79}    \\
$1920\pm15$ &20   &7 &\cite{brown79, baker79}     \\
$1950\pm15$ &--& --    &   \\
$1980\pm15$ &20  &7 &  \cite{brown79, baker79}  \\
$2010\pm15$ &20  &7 & \cite{brown79, baker79}   \\
$2040\pm15$ &--    &-- &   \\
$2070\pm15$ &20  &7 &  \cite{brown79, baker79}   \\
\hline
\end{tabular}
\end{center}
\end{table}

\begin{table}[htpb]
\caption{Statistics for single-energy fits for $\pi^- p\rightarrow K^0\Lambda$.}
\begin{center}
\begin{tabular}{|c|c|c|c|c|c|}
\hline
$W$(MeV) &${\rm d\sigma}/{\rm d\Omega}$ & $ P$ & $ P{\rm d\sigma}/{\rm d\Omega}$& $\beta$  & References\\ \hline
$1618\pm15$&25  &5&10 &--    &  \cite{knasel75, baker78} \\
$1648\pm15$&30  &10&10 &--    &  \cite{knasel75, baker78} \\
$1678\pm15$&170&10&80 &-- &   \cite{knasel75, baker78}\\
$1708\pm15$&90 &  10&40 &-- &   \cite{knasel75, baker78}\\
$1738\pm15$&30 &  14&10 &-- &   \cite{knasel75, baker78}\\
$1768\pm15$&10 &  14&--  &--&   \cite{baker78}\\
$1798\pm15$&10 &  14&--  & --&   \cite{baker78}\\
$1828\pm15$&10 &  14&-- &--&   \cite{baker78}\\
$1858\pm15$&10 &  14&-- &11&    \cite{baker78, bell83} \\
$1888\pm15$&20 &  20&-- &-- &    \cite{saxon80}\\
$1918\pm15$&33 &  20&11&--&    \cite{saxon80, dahl67}\\
$1948\pm15$&20 &  20&--  &9  &   \cite{saxon80, bell83}\\
$1978\pm15$&33 &  20&11 &--&   \cite{saxon80, dahl67}\\
$2008\pm15$&19 &  20&-- &--&   \cite{saxon80}\\
$2038\pm15$&33 &  19&11&10   &   \cite{saxon80, dahl67, bell83}\\
$2068\pm15$&20 &  18&-- &11&     \cite{saxon80, bell83}\\

\hline
\end{tabular}
\end{center}
\end{table}


   From Eqs.\ 1 to 4 it is clear that the amplitudes $f$ and $g$ can be multiplied by an arbitrary phase factor without changing the corresponding observables. This feature is referred to as the over-all phase ambiguity.
      For $\pi^-p\rightarrow\eta n$, the $S_{11}$ amplitudes below $K\Lambda$ threshold were held fixed at the values taken from the GWU solution (SP06) \cite{arndt06}. This constraint also removed the over-all phase ambiguity for the $\eta n$ amplitudes below 1.6 GeV. At higher energies, the phase ambiguity for $\eta n$ amplitudes was resolved by requiring the $G_{17}$ amplitude to have the same phase as the $G_{17}$ elastic amplitude. For $\pi^-p\rightarrow K^0\Lambda$, plots of $|T|^2$ vs. $W$ were made for all the contributing partial waves. The plot for the $S_{11}$ amplitude (Fig.\ 1) suggested a resonant behavior near 1.65 GeV where there is the well-established $S_{11}$(1650) resonance. The over-all phase problem for $K\Lambda$ amplitudes was thus resolved by rotating the amplitudes by a phase angle such that the rotated $S_{11}$ amplitude had a resonant phase consistent with our prior determinations of the $S_{11}$(1650) mass and width. 
      
\begin{figure}[H]
 \scalebox{0.6}{\includegraphics{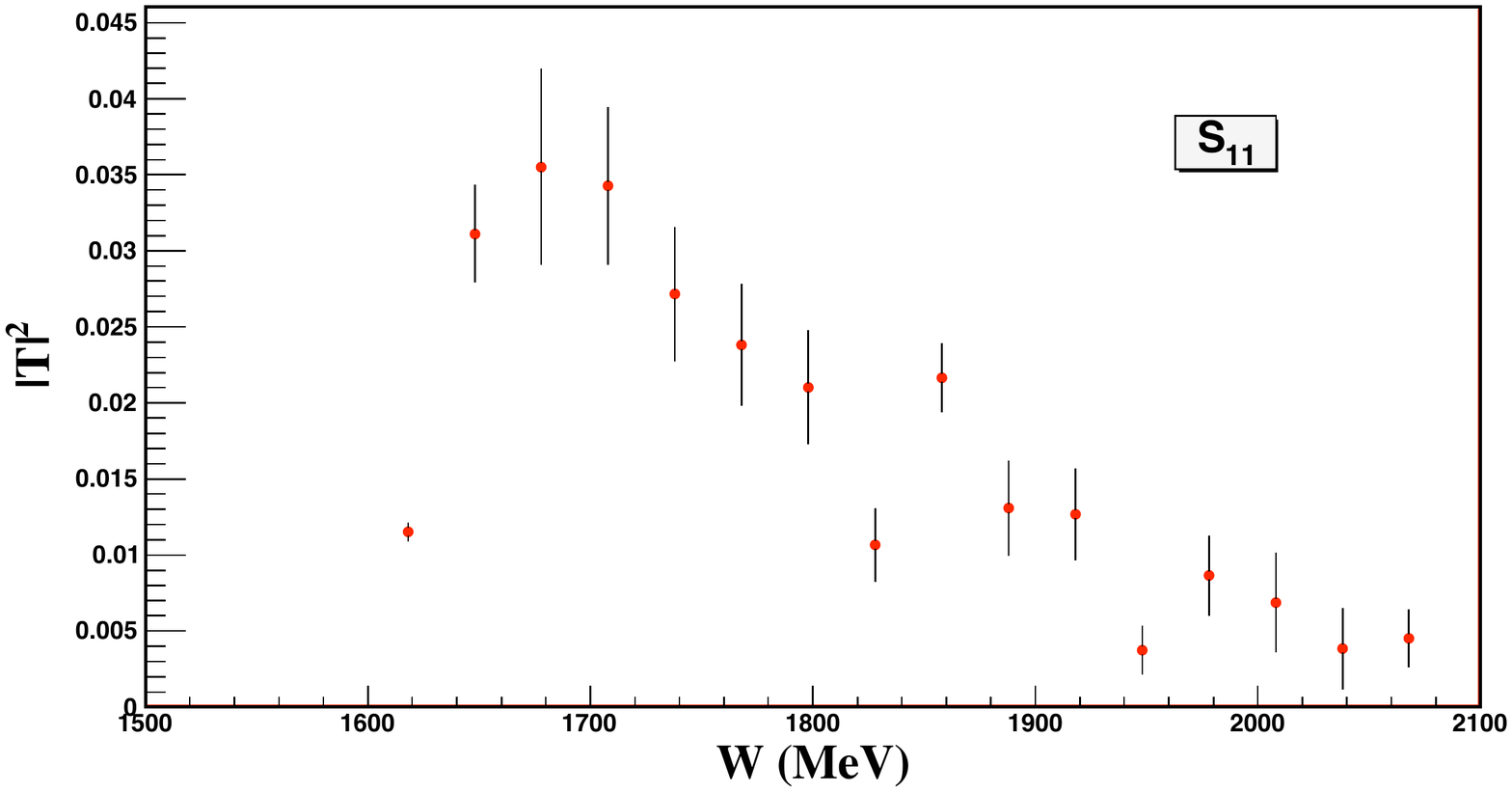}}  
 \caption{$|T|^2 $ vs. $W$ for the $\pi^- p\rightarrow K^0\Lambda$ $S_{11}$ amplitude.}
 \end{figure}
  
   In our initial fits, the single-energy solutions described the observables extremely well but with somewhat noisy amplitudes. These amplitudes were incorporated into a  global multichannel energy-dependent fit that yielded energy-dependent amplitudes consistent with two-body $S$-matrix unitarity. Details of the multichannel analysis will be presented in a separate publication \cite{manoj12}. 
  The initial energy-dependent amplitudes failed to reproduce the $\pi^{-}p\rightarrow\eta n$ and $\pi^{-}p\rightarrow K^0\Lambda$ observables satisfactorily so we iterated the single-energy fits. 
  
  Initially, for $K\Lambda$, only the $S_{11}$ amplitude was fitted well with the energy-dependent fit so in the second round of single-energy fits, the $S_{11}$ amplitude was held fixed at the energy-dependent values while the other partial-wave amplitudes were varied. The resulting constrained single-energy fits  still gave a very good description of the observables so we used this solution in the subsequent global energy-dependent fits. This time around the $P_{11}$ amplitude was fitted well. In the next round of single-energy fits, both the $S_{11}$ and $P_{11}$ amplitudes were held fixed at their energy-dependent values while the other amplitudes were varied. 
  
  Similarly,  for the second iteration of the $\eta N$ analysis, the $S_{11}$ and $P_{11}$ amplitudes were held fixed at their respective energy-dependent values while the other amplitudes were varied. In the final iteration, the  $S_{11}$, $P_{11}$ and $P_{13}$ amplitudes were held fixed, while the other amplitudes were varied. 

Our initial fits indicated that the $D_{13}$ amplitudes were not needed for either the $K\Lambda$ or the $\eta N$ fits. Thus the $D_{13}$ amplitudes were not included in our final single-energy solutions. This is consistent with the prior work that shows the inelasticity in $D_{13}$ is saturated by $\pi\pi N$ channels \cite{manley92}.\\

\section{\emph{\bf  RESULTS AND DISCUSSION }}
The final single-energy fits resulted in a fairly smooth set of partial-wave  amplitudes within the energy range of our analysis. Tables III and IV list the real and imaginary parts of the amplitudes tabulated against the central bin energies. The values in these tables represent the final single-energy solutions that were used as input into our subsequent global energy-dependent fits for given partial waves. 

\begin{table*}[htpb]
\caption{ Partial-wave amplitudes for $\pi N\rightarrow \eta N$.}
\begin{center}
\begin{tabular}{c|r|r|r|r|r|r}
\hline\hline
{$W$}  &  \multicolumn {2}{|c|}{$S_{11}$}  &   \multicolumn {2}{c}{$P_{11}$} &   \multicolumn {2}{|c}{$P_{13}$}\\\hline 
 (MeV)          &     \multicolumn {1}{|c|}{Re($S_{11}$)}   &  \multicolumn {1}{|c|}{Im($S_{11}$)}   & \multicolumn {1}{|c|}{Re($P_{11}$)}    &    \multicolumn {1}{|c|}{Im($P_{11}$)}&   \multicolumn {1}{|c|}{Re($P_{13}$)}   &  \multicolumn {1}{c}{Im($P_{13}$)}\\\hline
1530             &   &   &$-0.018\pm 0.025$    & $0.014\pm 0.025$     & $-0.000 \pm 0.063$   &$-0.004\pm 0.005$\\ 
1560             &   &  &$-0.043\pm 0.028$    & $-0.020\pm 0.022$    & $-0.005 \pm 0.040$    &$0.004\pm 0.023$\\
1590             &   &  &$-0.063\pm 0.018$    & $0.060\pm 0.016$   &  $-0.106 \pm 0.012$  &$-0.065\pm 0.019$\\     
1620             &  $-0.200 \pm 0.033$ &$0.156\pm 0.040$   &$-0.125\pm 0.036$    & $-0.025\pm 0.037$   &  $-0.024 \pm 0.017$   &$-0.004\pm 0.036$ \\   
1650             & $-0.218 \pm 0.061$  &$0.130\pm 0.057$   &$-0.122\pm 0.060$    & $-0.026\pm 0.058$   & $-0.002 \pm 0.021$    &$-0.000\pm 0.032$  \\ 
1680             & $-0.051 \pm 0.039$  &$-0.205\pm 0.038$  &$0.051\pm 0.039$     & $-0.122\pm 0.030$    & $-0.008 \pm 0.027$    &$-0.023\pm 0.030$ \\ 
1710             & $-0.042 \pm 0.046$  &$-0.225\pm 0.037$  &$0.039\pm 0.049$     & $-0.197\pm 0.036$    & $0.058 \pm 0.037$     &$0.013\pm 0.041$  \\ 
1770             & $-0.033 \pm 0.043$  &$-0.197\pm 0.023$  &$0.044\pm 0.046$     & $-0.193\pm 0.030$    & $0.021 \pm 0.024$     &$0.028\pm 0.028$\\ 
1830             & $0.196 \pm 0.035$   &$0.143\pm 0.041$    &$0.120\pm 0.038$    & $0.144\pm 0.034$     &  $0.047 \pm 0.022$   &$0.020\pm 0.029$\\ 
1860             & $0.181 \pm 0.019$   &$0.160\pm 0.023$    &$0.091\pm 0.022$    & $0.207\pm 0.020$     & $ 0.074 \pm 0.026$   &$0.003\pm 0.019$\\
1890             & $0.192 \pm 0.025$   &$0.140\pm 0.031$    &$0.077\pm 0.025$    & $0.169\pm 0.024$     & $0.063 \pm 0.016$    &$0.003\pm 0.017$\\ 
1920             & $0.158 \pm 0.036$   &$0.131\pm 0.041$    &$0.056\pm 0.033$    & $0.153\pm 0.032$     & $0.066 \pm 0.026$    &$0.001\pm 0.017$\\ 
1980             & $0.138 \pm 0.054$   &$0.143\pm 0.057$   &$0.027\pm 0.042$    & $0.114\pm 0.042$     & $0.047 \pm 0.041$     &$0.082\pm 0.041$\\ 
2010             & $0.112 \pm 0.059$   &$0.128\pm 0.062$   &$0.019\pm 0.043$    & $0.063\pm 0.035$     & $0.072 \pm 0.039$     &$0.052\pm 0.035$ \\ 
2070             &$0.089 \pm 0.047$    &$0.156\pm 0.048$   &$-0.008\pm 0.030$   & $0.092\pm 0.025$     & $0.084 \pm 0.033$     &$0.098\pm 0.027$ \\ 
\hline
\hline
\end{tabular}
\end{center}
\end{table*}

  \begin{table*}[htpb]
\addtocounter{table}{-1}
\caption{ Continued.}
\label{Hadronic}
\begin{center} 
\begin{tabular}{c|r|r|r|r|r|r}
\hline
\hline 
$W$   &   \multicolumn {2}{|c|}{$D_{15}$} &   \multicolumn {2}{|c|}{$F_{15}$}  &   \multicolumn {2}{c}{$G_{17}$}\\\hline
(MeV)            & \multicolumn {1}{|c|}{Re($D_{15}$)}   & \multicolumn {1}{|c|}{Im($D_{15}$)} & \multicolumn {1}{|c|}{Re($F_{15}$)} & \multicolumn {1}{|c|}{Im($F_{15}$)} &\multicolumn {1}{|c|}{Re($G_{17}$)} &   \multicolumn {1}{c}{Im($G_{17}$)}\\\hline
1530             & $ 0.007 \pm 0.015$  &$0.013\pm 0.002$&   &  &    & \\ 
1560             & $ 0.002 \pm 0.022$  &$0.035\pm 0.008$& &  &     & \\
1590             &  $ 0.041 \pm 0.011$  &$ 0.086\pm 0.003$ &  &     &   &\\     
1620             &  $-0.031 \pm 0.016$  &$0.032\pm 0.034$&  $ 0.042 \pm 0.017$  &$-0.023\pm 0.034$ &   & \\   
1650             & $-0.040 \pm 0.030$  &$ 0.005\pm 0.024$& $ 0.044 \pm 0.030$  &$ 0.020\pm 0.023$ &    & \\ 
1680             & $-0.054 \pm 0.030$  &$0.024\pm 0.037$& $0.033\pm 0.030$  &$ 0.014\pm 0.037$ &    & \\ 
1710             & $-0.001 \pm 0.029$  &$-0.023\pm 0.030$& $-0.045\pm 0.026$  &$ 0.056\pm 0.032$ &$-0.035\pm 0.024$       & $0.031\pm 0.024$\\ 
1770             & $0.039 \pm 0.015$  &$0.007\pm 0.021$& $-0.094 \pm 0.012$  &$-0.000\pm 0.026$ &$-0.041\pm 0.021$       & $-0.041\pm 0.021$\\ 
1830             & $-0.011 \pm 0.021$  &$-0.001\pm 0.021$& $-0.042 \pm 0.027$  &$-0.052\pm 0.024$ &$-0.027\pm 0.016$       & $-0.038\pm 0.016$\\ 
1860             & $-0.005 \pm 0.011$  &$0.032\pm 0.009$ & $-0.085 \pm 0.010$  &$-0.047\pm 0.012$ &$-0.037\pm 0.010$    & $-0.017\pm 0.010$\\ 
1890             & $-0.031 \pm 0.010$  &$0.018\pm 0.014$& $-0.044 \pm 0.013$  &$-0.050\pm 0.012$ &$-0.060\pm 0.009$     & $-0.029\pm 0.009$\\ 
1920             & $-0.047\pm 0.011$  &$0.023\pm 0.013$& $-0.046 \pm 0.017$  &$-0.054\pm 0.015$ &$-0.042\pm 0.010$     & $-0.030\pm 0.010$\\ 
1980             & $-0.024 \pm 0.026$  &$-0.011\pm 0.023$& $-0.034 \pm 0.027$  &$-0.066\pm 0.020$ &$-0.054\pm 0.022$   & $-0.064\pm 0.022$\\ 
2010             & $0.016 \pm 0.032$  &$-0.009\pm 0.022$& $-0.027 \pm 0.034$  &$-0.045\pm 0.014$ &$-0.021\pm 0.026$    & $-0.060\pm 0.026$\\ 
2070             & $0.040 \pm 0.019$  &$-0.011\pm 0.012$& $-0.035 \pm 0.024$  &$-0.076\pm 0.011$ &$-0.024\pm 0.019$    & $-0.071\pm 0.019$\\ 
\hline
\hline
\end{tabular}
\end{center}
\end{table*}

\begin{table*}[htpb]
\caption{Partial-wave amplitudes for $\pi N\rightarrow K\Lambda$.}
\begin{center}
\begin{tabular}{c|r|r|r|r|r|r}
\hline\hline
$W$   &   \multicolumn {2}{|c|}{$S_{11}$}  &   \multicolumn {2}{c}{$P_{11}$} &   \multicolumn {2}{|c}{$P_{13}$}\\\hline
(MeV)            & \multicolumn {1}{|c|}{Re($S_{11}$)} & \multicolumn {1}{|c|}{Im($S_{11}$)} &\multicolumn {1}{|c|}{Re($P_{11}$)} & \multicolumn {1}{|c|}{Im($P_{11}$)}& \multicolumn {1}{|c|}{Re($P_{13}$)} & \multicolumn {1}{c}{Im($P_{13}$)}\\\hline
1618             & $-0.065 \pm 0.003$   &$-0.085\pm 0.003$ &$0.047\pm 0.007$    & $-0.067\pm 0.008$ & $0.006 \pm 0.005$  &$0.006\pm 0.015$\\ 
1648             & $-0.052 \pm 0.003$   &$-0.169\pm 0.010$ &$0.073\pm 0.007$    & $-0.096\pm 0.003$ & $-0.008 \pm 0.006$  &$-0.003\pm 0.005$\\ 
1678             & $0.028 \pm 0.003$   &$-0.186\pm 0.017$ &$0.118\pm 0.004$     & $-0.093\pm 0.005$& $ -0.029 \pm 0.005$  &$-0.035\pm 0.003$\\
1708             &  $0.097 \pm 0.009$  &$-0.158\pm 0.015$ &$0.152\pm 0.011$    & $0.025\pm 0.016$ &  $-0.011 \pm 0.007$  &$ -0.061\pm 0.006$\\     
1738             &  $0.120 \pm 0.014$  &$-0.113\pm 0.013$ &$0.153\pm 0.017$    & $0.025\pm 0.029$ &  $-0.017 \pm 0.008$  &$-0.062\pm 0.008$ \\   
1768             & $0.129 \pm 0.014$   &$-0.085\pm 0.009$ &$0.123\pm 0.022$    & $0.082\pm 0.023$ & $ -0.002 \pm 0.007$  &$- 0.061\pm 0.006$  \\ 
1798             & $0.130 \pm 0.014$   &$-0.065\pm 0.007$ &$0.106\pm 0.025$    & $0.129\pm 0.023$ & $ -0.018 \pm 0.008$  &$-0.071\pm 0.010$ \\ 
1828             & $0.096 \pm 0.013$   &$-0.039\pm 0.005$ &$0.062\pm 0.022$    & $0.050\pm 0.014$  & $0.007 \pm 0.009$  &$-0.028\pm 0.009$  \\ 
1858             & $0.139 \pm 0.008$   &$-0.048\pm 0.003$ &$0.068\pm 0.010$    & $0.064\pm 0.013$  & $0.022 \pm 0.011$  &$-0.045\pm 0.006$ \\ 
1888             & $0.110 \pm 0.014$   &$-0.033\pm 0.004$ &$0.072\pm 0.016$    & $0.070\pm 0.020$  & $-0.002 \pm 0.012$  &$-0.080\pm 0.008$\\ 
1918             & $0.109 \pm 0.014$   &$-0.029\pm 0.004$ &$0.077\pm 0.024$    & $0.057\pm 0.020$  &  $0.043 \pm 0.018$  &$-0.099\pm 0.011$\\ 
1948             & $0.060 \pm 0.013$  &$-0.014\pm 0.003$ &$0.086\pm 0.016$    & $0.062\pm 0.013$  &  $0.026 \pm 0.010$  &$-0.037\pm 0.008$\\ 
1978             & $0.091 \pm 0.014$  &$-0.019\pm 0.003$ &$0.036\pm 0.018$    & $0.032\pm 0.010$ & $0.039 \pm 0.019$  &$-0.002\pm 0.011$\\
2008             & $0.081 \pm 0.020$  &$-0.016\pm 0.004$ &$0.063\pm 0.018$    & $0.023\pm 0.021$ & $0.028 \pm 0.015$  &$-0.007\pm 0.013$\\ 
2038             & $0.061 \pm 0.022$  &$-0.011\pm 0.004$ &$0.072\pm 0.012$    & $0.034\pm 0.027$  & $0.019 \pm 0.014$  &$0.004\pm 0.010$\\ 
2068             & $0.066 \pm 0.014$  &$-0.011\pm 0.002$ &$0.071\pm 0.011$    & $0.032\pm 0.019$  & $0.031 \pm 0.008$  &$-0.017\pm 0.013$ \\ 
\hline
\hline
\end{tabular}
\end{center}
\end{table*}

  \begin{table*}[htpb]
\addtocounter{table}{-1}
\caption{ Continued.}
\label{Hadronic}
\begin{center} 
\begin{tabular}{c|r|r|r|r|r|r}
\hline
\hline 
$W$   &   \multicolumn {2}{|c|}{$D_{15}$}  &   \multicolumn {2}{c}{$F_{15}$}&   \multicolumn {2}{|c}{$F_{17}$}\\\hline
(MeV)           &\multicolumn {1}{|c|}{Re($D_{15}$)}& \multicolumn {1}{|c|}{Im($D_{15}$)} &\multicolumn {1}{|c|}{Re($F_{15}$)} &  \multicolumn {1}{c}{ Im($F_{15}$)}& \multicolumn {1}{|c|}{Re($F_{17}$)} & \multicolumn {1}{c}{Im($F_{17}$)}\\\hline

1618             & $-0.014 \pm 0.006$   &$0.032\pm 0.005$ &$-0.010\pm 0.009$    & $-0.011\pm 0.004$    &    &   \\
1648             & $-0.013 \pm 0.007$  &$0.006\pm 0.008$ &$-0.009\pm 0.009$    & $-0.012\pm 0.008$   &    &   \\ 
1678             & $-0.006 \pm 0.006$  &$-0.001\pm 0.005$ &$0.016\pm 0.009$    & $-0.011\pm 0.007$   &    &   \\
1708            &  $-0.009 \pm 0.007$  &$0.003\pm 0.006$ &$0.025\pm 0.011$    & $-0.008\pm 0.016$    &    &    \\     
1738            &  $-0.023 \pm 0.008$  &$-0.018\pm 0.011$ &$0.027\pm 0.012$    & $-0.007\pm 0.017$   &    &   \\   
1768            & $-0.006 \pm 0.010$  &$-0.024\pm 0.011$ &$0.010\pm 0.010$    & $0.003\pm 0.009$     &    &   \\ 
1798            & $-0.009 \pm 0.012$  &$-0.033\pm 0.010$ &$0.021\pm 0.017$    & $0.003\pm 0.014$     &    &   \\ 
1828            & $-0.001 \pm 0.017$  &$-0.026\pm 0.016$ &$0.006\pm 0.016$    & $0.014\pm 0.016$     &    &   \\ 
1858            & $-0.018\pm 0.009$   &$-0.059\pm 0.006$ &$0.004\pm 0.008$    & $-0.024\pm 0.008$    &$0.005\pm0.008$   &$-0.014\pm0.006$\\ 
1888            & $0.007 \pm 0.008$  &$-0.067\pm 0.007$ &$0.011\pm 0.008$    & $-0.009\pm 0.009$     &$-0009\pm0.008$   &$-0.003\pm0.008$\\ 
1918            & $0.023 \pm 0.008$  &$-0.086\pm 0.016$ &$0.009\pm 0.007$    & $-0.011\pm 0.007$     &$-0.010\pm0.007$  &$-0.001\pm0.005$\\ 
1948            & $-0.022 \pm 0.010$  &$-0.080\pm 0.007$ &$-0.003\pm 0.014$    & $-0.024\pm 0.008$  &$-0.003\pm0.006$  &$-0.008\pm0.005$\\ 
1978            & $-0.022 \pm 0.012$  &$-0.073\pm 0.007$ &$0.007\pm 0.007$    & $-0.007\pm 0.008$   &$-0.005\pm0.005$   &$-0.008\pm0.004$\\ 
2008            & $0.011 \pm 0.013$  &$-0.066\pm 0.007$ &$-0.014\pm 0.009$    & $-0.004\pm 0.008$   &$0.010\pm0.008$    &$-0.011\pm0.008$\\ 
2038            & $0.003 \pm 0.009$  &$-0.060\pm 0.006$ &$-0.022\pm 0.006$    & $-0.007\pm 0.008$    &$0.009\pm0.006$    &$-0.001\pm0.006$\\ 
2068            & $0.004 \pm 0.005$  &$-0.050\pm 0.006$ &$-0.007\pm 0.007$    & $0.011\pm 0.008$    &$0.018\pm0.005$    &$-0.014\pm0.006$\\ 
\hline
\hline
\end{tabular}
\end{center}
\end{table*}

 \begin{table*}[htpb]
\addtocounter{table}{-1}
\caption{ Continued.}
\label{Hadronic}
\begin{center} 
\begin{tabular}{c|r|r|r|r|r|r}
\hline
\hline 
$W$   &   \multicolumn {2}{|c|}{$G_{17}$}  &   \multicolumn {2}{c}{$G_{19}$}&   \multicolumn {2}{|c}{$H_{19}$}\\\hline
(MeV)           &\multicolumn {1}{|c|}{Re($G_{17}$)}& \multicolumn {1}{|c|}{Im($G_{17}$)} &\multicolumn {1}{|c|}{Re($G_{19}$)} &  \multicolumn {1}{c}{ Im($G_{19}$)}& \multicolumn {1}{|c|}{Re($H_{19}$)} & \multicolumn {1}{c}{Im($H_{19}$)}\\\hline

1858            & $-0.002\pm 0.007$   &$0.012\pm 0.006$ &$-0.009\pm 0.007$    & $-0.003\pm 0.005$   &$0.003\pm0.005$   &$-0.011\pm0.006$\\ 
1888            & $0.006 \pm 0.008$  &$-0.001\pm 0.005$ &$-0.003\pm 0.006$    & $0.001\pm 0.007$    &$-0.006\pm0.006$  &$-0.005\pm0.005$\\
1918            & $0.013 \pm 0.005$  &$0.003\pm 0.004$ &$-0.016\pm 0.005$    & $-0.002\pm 0.005$    &$-0.007\pm0.005$  &$-0.003\pm0.004$\\ 
1948            & $0.008 \pm 0.005$  &$0.004\pm 0.004$ &$-0.013\pm 0.004$    & $-0.016\pm 0.005$    &$-0.007\pm0.005$  &$0.009\pm0.004$\\ 
1978            & $-0.011 \pm 0.007$  &$0.023\pm 0.004$ &$-0.000\pm 0.004$    & $-0.009\pm 0.005$   &$0.028\pm0.005$   &$-0.003\pm0.006$\\ 
2008            & $-0.002 \pm 0.012$  &$0.013\pm 0.008$ &$-0.010\pm 0.007$    & $-0.008\pm 0.008$   &$0.021\pm0.009$   &$-0.018\pm0.008$\\ 
2038            & $-0.005 \pm 0.007$  &$0.009\pm 0.006$ &$0.007\pm 0.005$    & $-0.028\pm 0.006$     &$0.005\pm0.005$   &$-0.002\pm0.005$\\ 
2068            & $-0.005 \pm 0.005$  &$-0.007\pm 0.004$ &$0.001\pm 0.006$    & $0.000\pm 0.006$    &$0.010\pm0.004$   &$0.003\pm0.003$\\ 
\hline
\hline
\end{tabular}
\end{center}
\end{table*}

\newpage
Figure 2 shows representative final single-energy fits for the $\pi^- p \rightarrow \eta n$ differential cross section for bins centered at $W$ = 1530,  1590, 1680, 1770, 1890, and 2010 MeV. The first three panels of Fig.~2 reveal an inconsistency in the data over the small variation in c.m.\ energy within the bin. Note that in these panels, there are data from the same references (Brown {\it et al.\ }\cite{brown79} and Debeham {\it et al.\ }\cite{debeham75}) at two  or more slightly different energies. Our single-energy analysis gives a weighted average fit to the data sets. Figure 3 similarly shows final single-energy fits for the reaction $\pi^{-}p\rightarrow K^0\Lambda$ differential cross section for bins centered at $W$ = 1648, 1738, 1828, 1918, 1978, and 2038 MeV.
Figure 4 shows final single-energy fits for the spin rotation parameter for $\pi^{-}p\rightarrow K^0\Lambda$. In general, our final results are in very good agreement with the available data.

Figure 5 shows our predictions for the integrated cross sections for the two inelastic reactions obtained using our energy-dependent amplitudes. In the threshold region for $\pi^-p\rightarrow \eta n$, the different data do not agree well with each other but our prediction is in excellent agreement with the latest and more precise data by Prakhov {\it et al.\ }\cite{prakhov05}. Figure 5 also shows the individual contributions from the dominant partial waves.  From this break-down, it is clear that the $S_{11}$ amplitude (red curve) dominates the peak region associated with the $S_{11}$(1535) resonance. However, the contribution from $S_{11}$ is small in the vicinity of the $S_{11}(1650)$. The next important partial wave is $P_{11}$ with a peak around 1700 MeV as shown by the brown curve. The contributions from other partial waves are small. Also one can see a small cusp effect on the blue curve (KSU prediction) near 1620 MeV that indicates the opening of the $K\Lambda$ channel.
For $\pi^-p\rightarrow K^0\Lambda$, the peak near 1700 MeV is described by almost equal contributions from the $S_{11}$ and $P_{11}$ partial waves, which suggests considerable $K\Lambda$ coupling to the $S_{11}(1650)$ and $P_{11} (1710)$ resonances. The $P_{13}$ partial wave (black curve) contributes to the small suggested peak near 1900 MeV. This feature is consistent with a prominent peak seen near 1900 MeV in the reaction $\gamma p \rightarrow K^+\Lambda$ \cite{aniso11}.

Early analyses of $\pi^-p\rightarrow \eta n$ were energy-dependent PWAs based on a simple assumption that the partial-wave amplitudes could be parameterized as either $T = T_B + T_R$ \cite{feltesse75} or $T = T_R$ without a background term \cite{baker79}. The 1975 analysis by Feltesse {\it et al.\ }\cite{feltesse75} used far fewer data than the 1979 analysis by Baker {\it et al.\ }\cite{baker79}, which included polarization data unlike the earlier analysis of Ref. \cite{feltesse75}. Both analyses violated $S$-matrix unitarity.
Our results for $\pi N\rightarrow \eta N$ differ significantly from those of Baker {\it et al.\ }\cite{baker79}. Firstly, partial waves above $G_{17}$ were not needed in the present analysis but that of Baker {\it et al.\ }included partial waves up to $H_{19}$. Also the $D_{13}$ amplitude was found to be negligible over the entire energy range in the present work. Secondly, the $S_{11}$ wave was poorly determined by Baker {\it et al.,\ }especially near the threshold region. This could be due to poor data but our prediction for the integrated cross section agrees satisfactorily with the precise and recent data from Prakhov {\it et al.\ }\cite{prakhov05}. This re-enforces the reliability of the $S_{11}$ amplitude from our analysis. The other partial waves where we disagree with Baker {\it et al.\ }are $P_{13}$ and $F_{15}$ at low energies. For the $P_{11}$ amplitude, both Ref.~\cite{baker79} and the present work find significant contributions near 1700 MeV. The more recent 1995 analysis by Batani\'c {\it et al.} \cite{batinic95} used a three-channel unitary approach to obtain partial-wave amplitudes for $\pi N\rightarrow \pi N$ and $\pi N\rightarrow \eta N$, and to predict the same for $\eta N\rightarrow \eta N$. There is a striking resemblance of the $S_{11}$ and $P_{11}$ amplitudes between our analysis and one of the solutions in 
Ref.~\cite{batinic95}. For higher partial waves, the differences are more pronounced. Also the analysis of Batani\'c {\it et al.\ }required the $D_{13}$ and $F_{17}$ amplitudes, which were not needed in the present work. Our analysis does better in describing the $\pi^-p\rightarrow \eta n$ differential cross section, especially at forward angles in the c.m.\ energy range 1650 to 2070 MeV, than either the 2008 EBAC analysis \cite{durand08} based on a dynamical coupled-channels (DCC) model or the 2011 analysis  \cite{julich12} by the J\"ulich-Athens group based on the J\"ulich DCC model. 

 For $\pi^-p\rightarrow K^0\Lambda$, the previous analyses \cite{knasel75,baker78,saxon80,bell83} were also energy-dependent PWAs based on the simple parameterization $T=T_B+T_R$, which violates $S$-matrix unitarity. The 1970s analyses of $\pi^-p\rightarrow K^0\Lambda$ by Knasel {\it et al.\ }\cite{knasel75}, Baker {\it et al.\ }\cite{baker79}, and Saxon {\it et al.\ }\cite{saxon80} used differential cross section, polarization and/or polarized cross section data, but no spin-rotation data. The 1983 analysis by Bell {\it et al.\ }\cite{bell83} included differential cross section and polarization data from their previous analyses \cite{baker78,saxon80}, plus spin-rotation data. Our results broadly agree with these previous analyses, especially that by Bell {\it et al.\ }\cite{bell83}. Their $S_{11}$ amplitude has a similar behavior to ours except for an opposite over-all sign. The main difference is $F_{15}$ is not required in Ref.~\cite{bell83} but is included in our work. The description of spin-rotation measurements by the present single-energy analysis is better in some cases than that by Bell {\it et al.\ }\cite{bell83} and is as good as the recent work by the Bonn-Gatchina group \cite{sarantsev11}. Also the contributions we find for the leading partial waves (see Fig.~5) agree very well with the analysis by the Bonn-Gatchina group. 
\begin{figure*}
\vspace{-45mm}
\scalebox{0.6}{\includegraphics{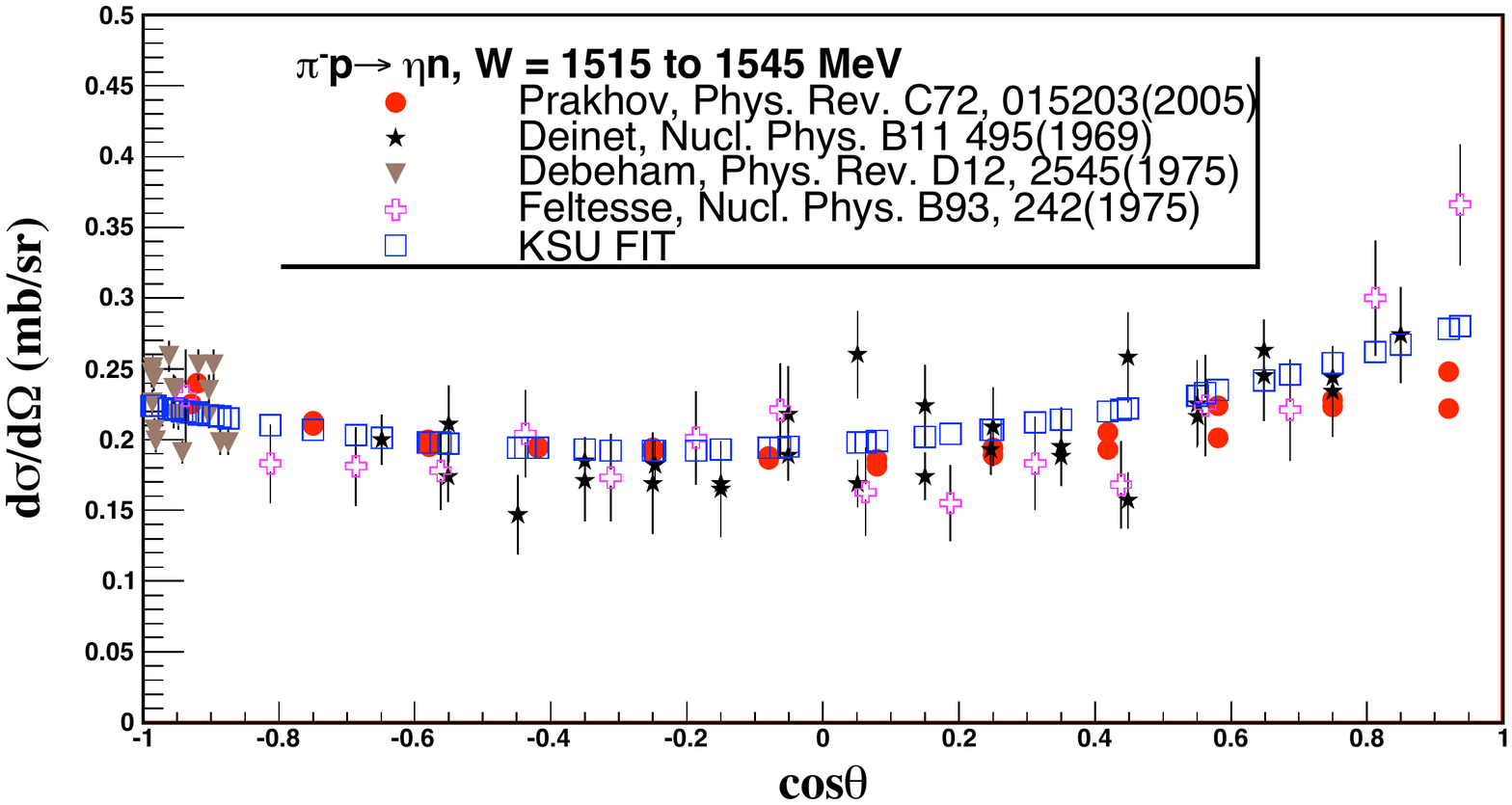}}
\vspace{-30mm}
\vspace{-33mm}
\scalebox{0.6}{\includegraphics{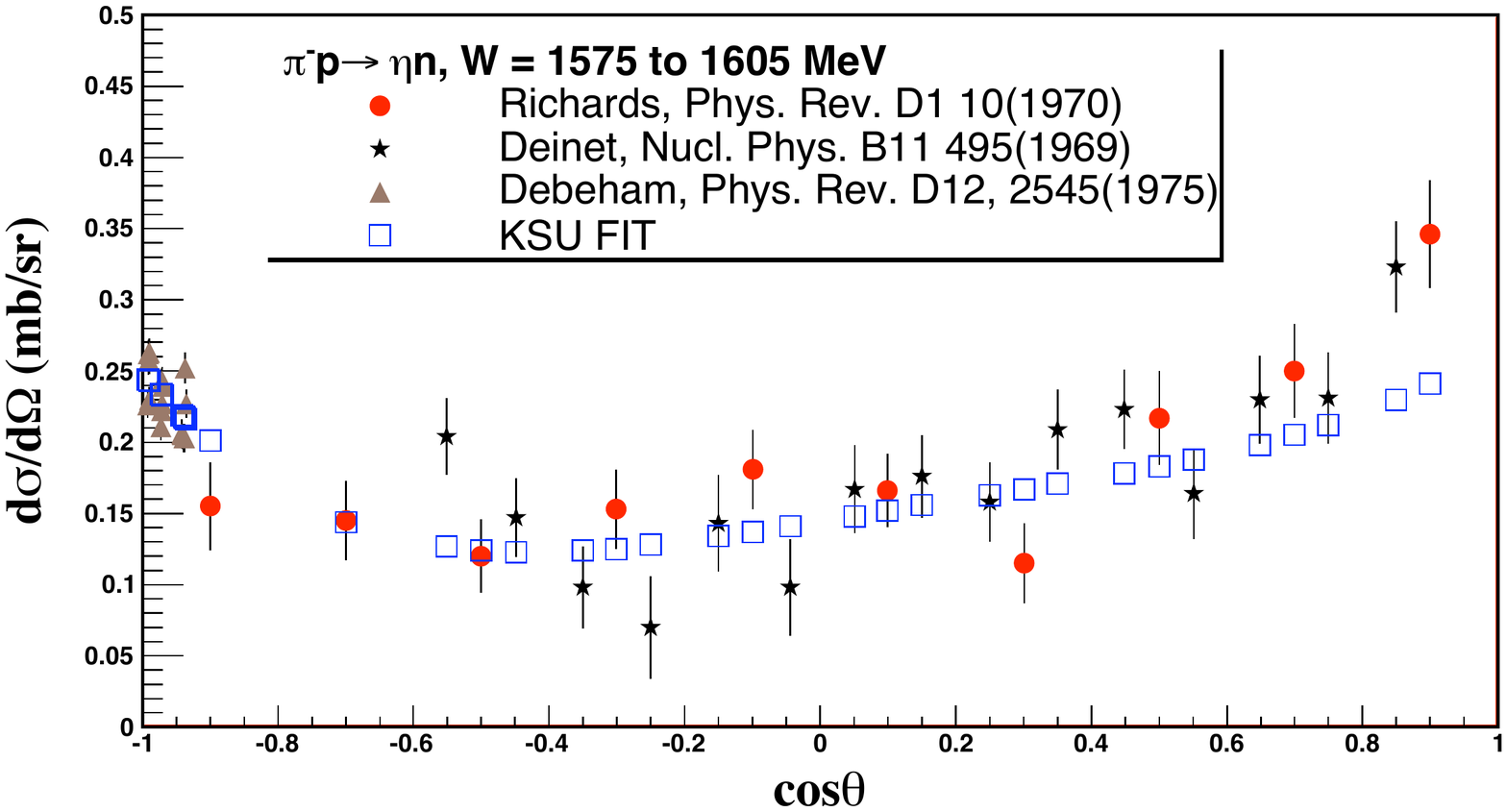}}
\vspace{-48mm}
\vspace{-15mm}
\scalebox{0.6}{\includegraphics{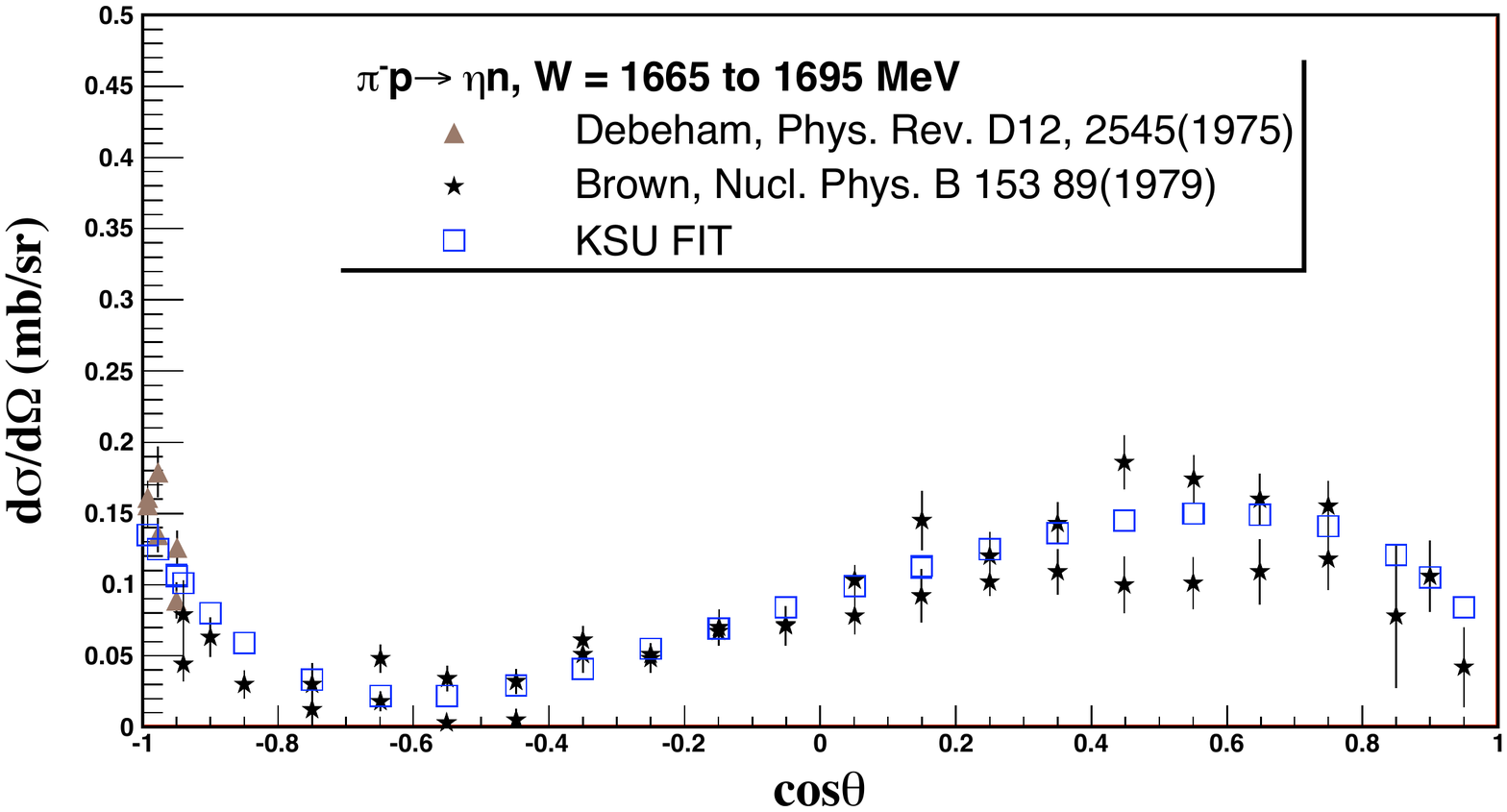}}
\vspace{-15mm}
\caption{Single-energy fit results for $\pi^- p\rightarrow \eta n$.}
\end{figure*}

\begin{figure*}
\addtocounter{figure}{-1}
\vspace{-45mm}
\scalebox{0.6}{\includegraphics{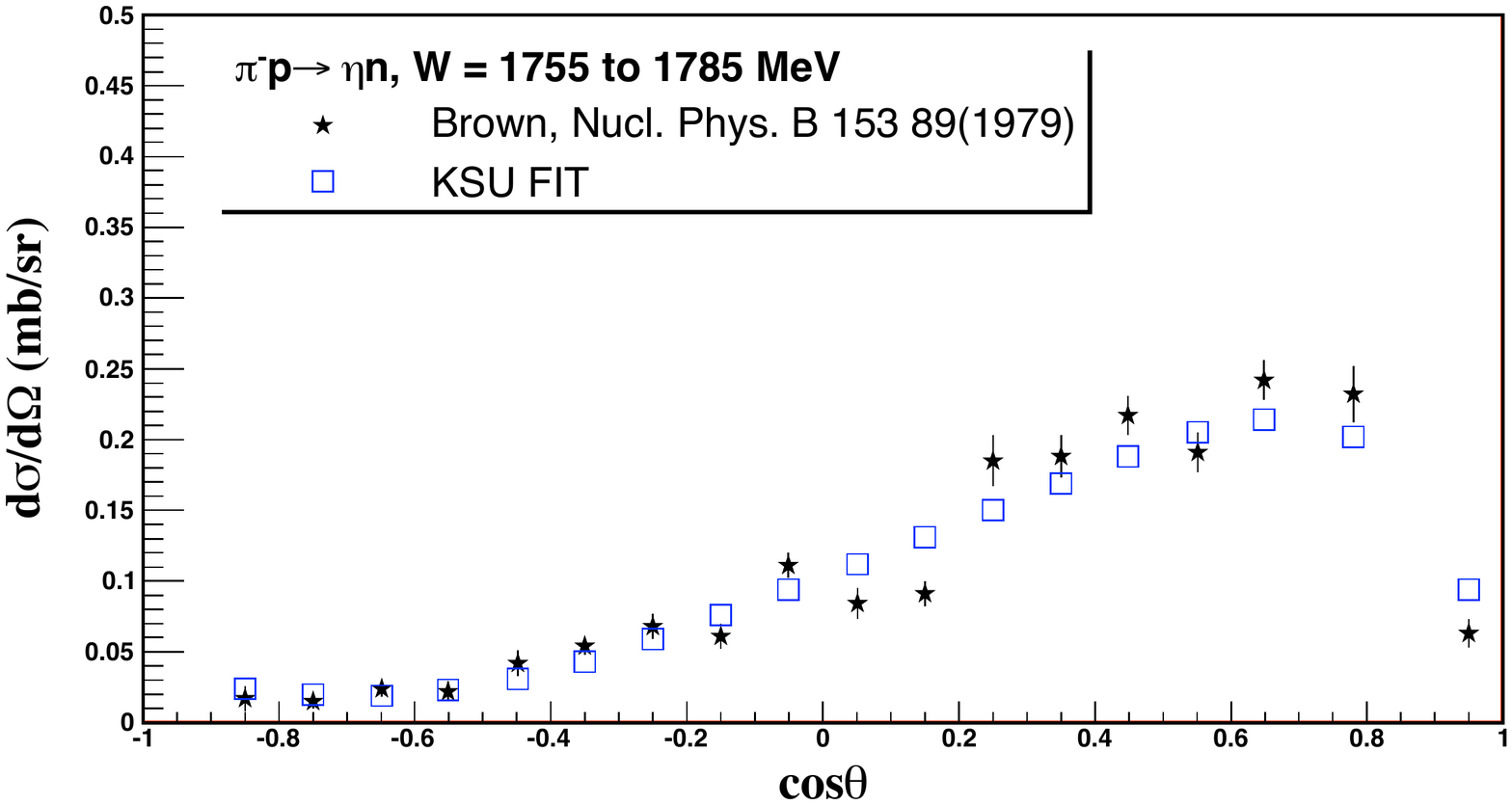}}
\vspace{-31mm}
\vspace{-33mm}
\scalebox{0.6}{\includegraphics{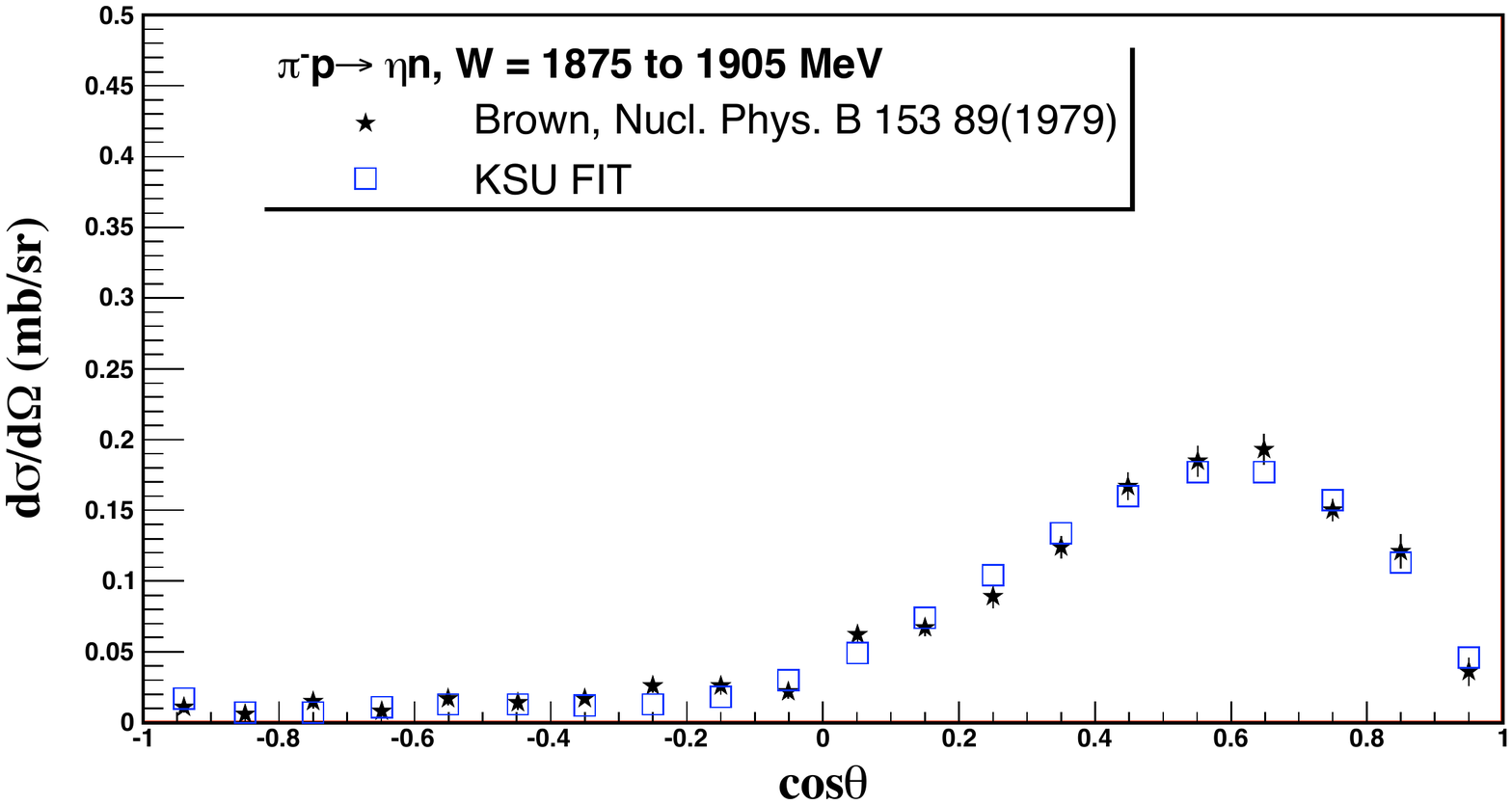}}
\vspace{-49mm}
\vspace{-15mm}
\scalebox{0.6}{\includegraphics{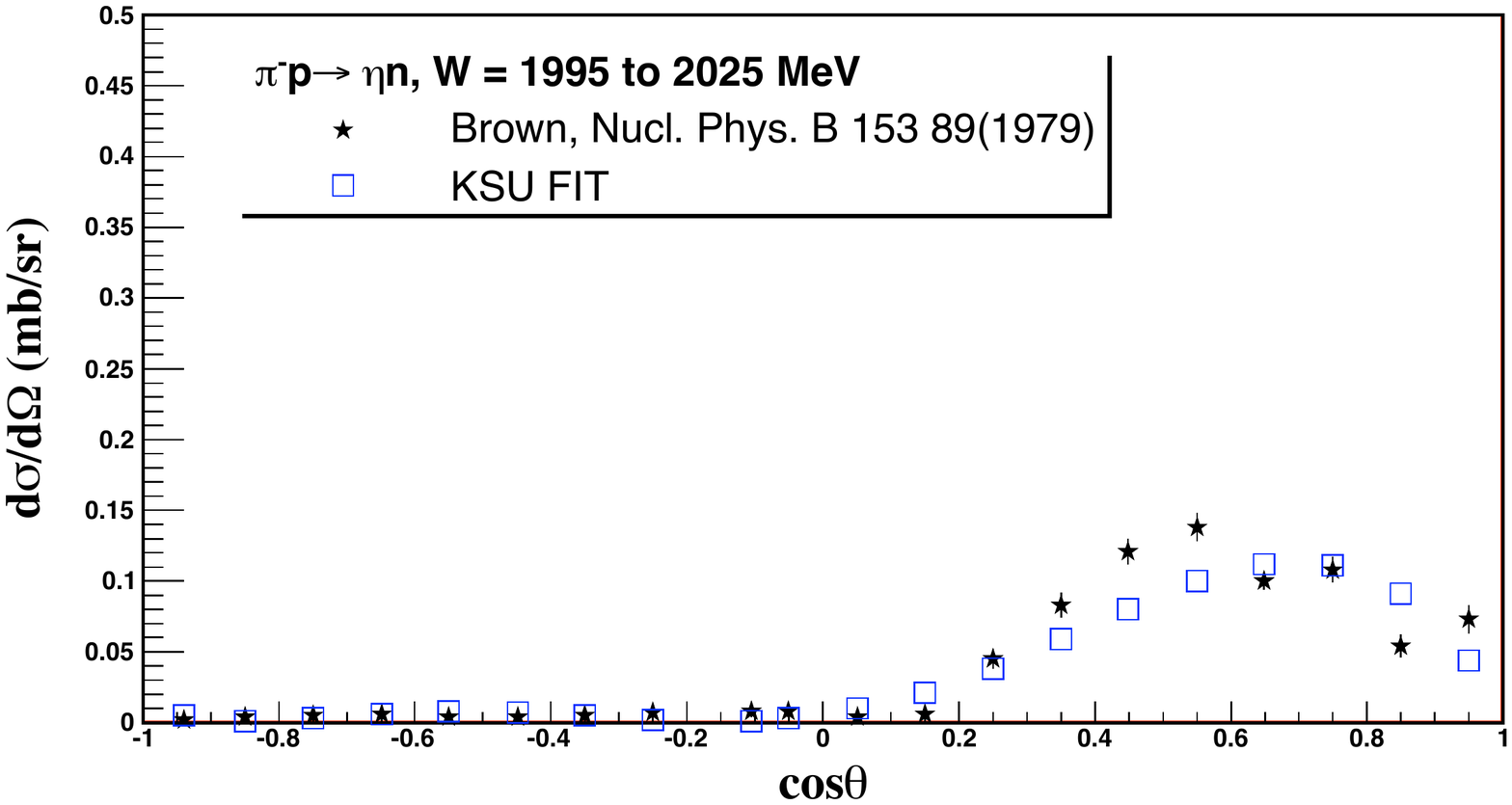}}
\vspace{-15mm}
\caption{(Cont'd)}
\label{fig:Amplitudes}
\end{figure*}

\begin{figure*}[htpb]
\vspace{-45mm}
\scalebox{0.6}{\includegraphics{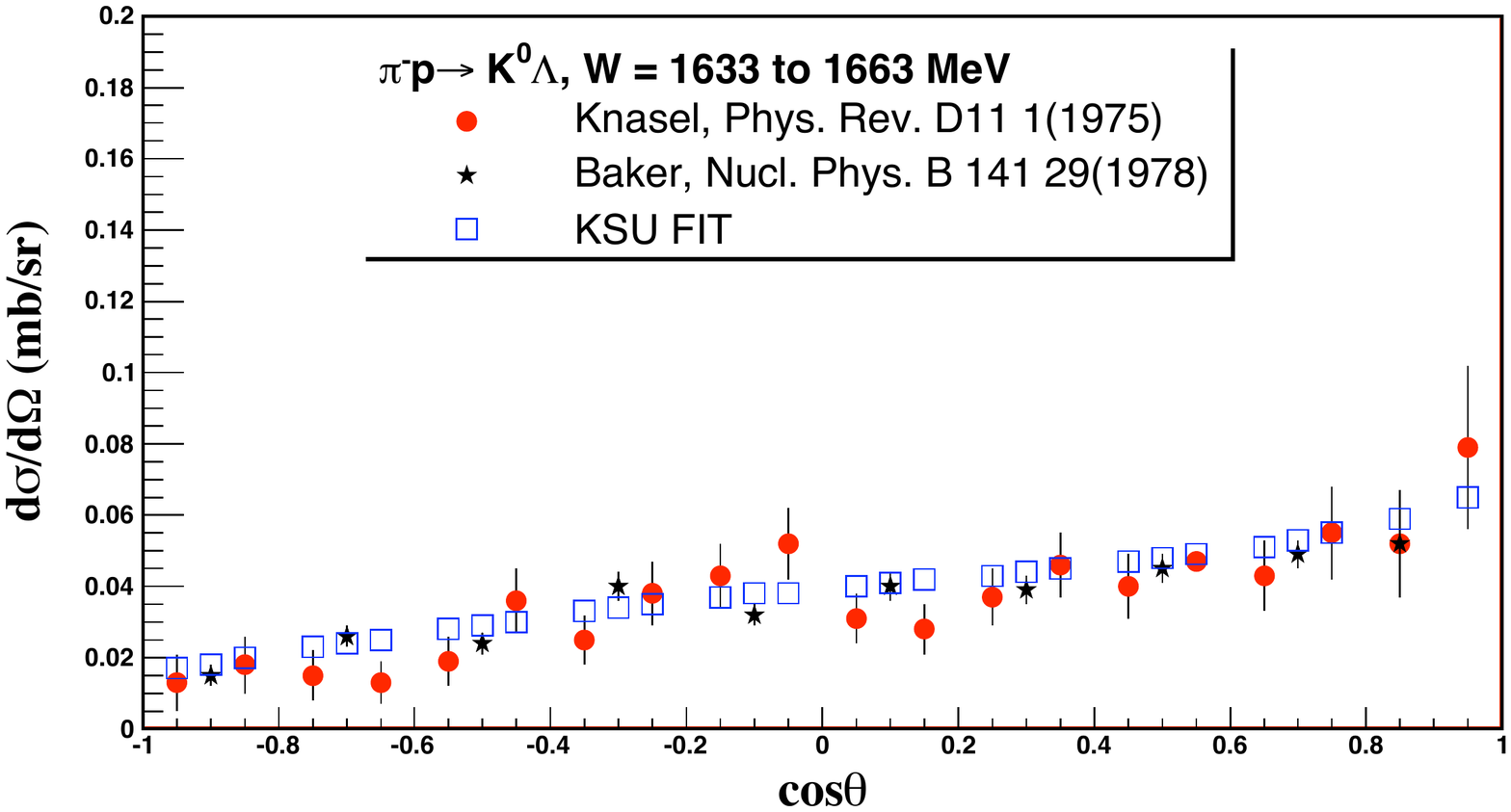}}
\vspace{-31mm}
\vspace{-33mm}
\scalebox{0.6}{\includegraphics{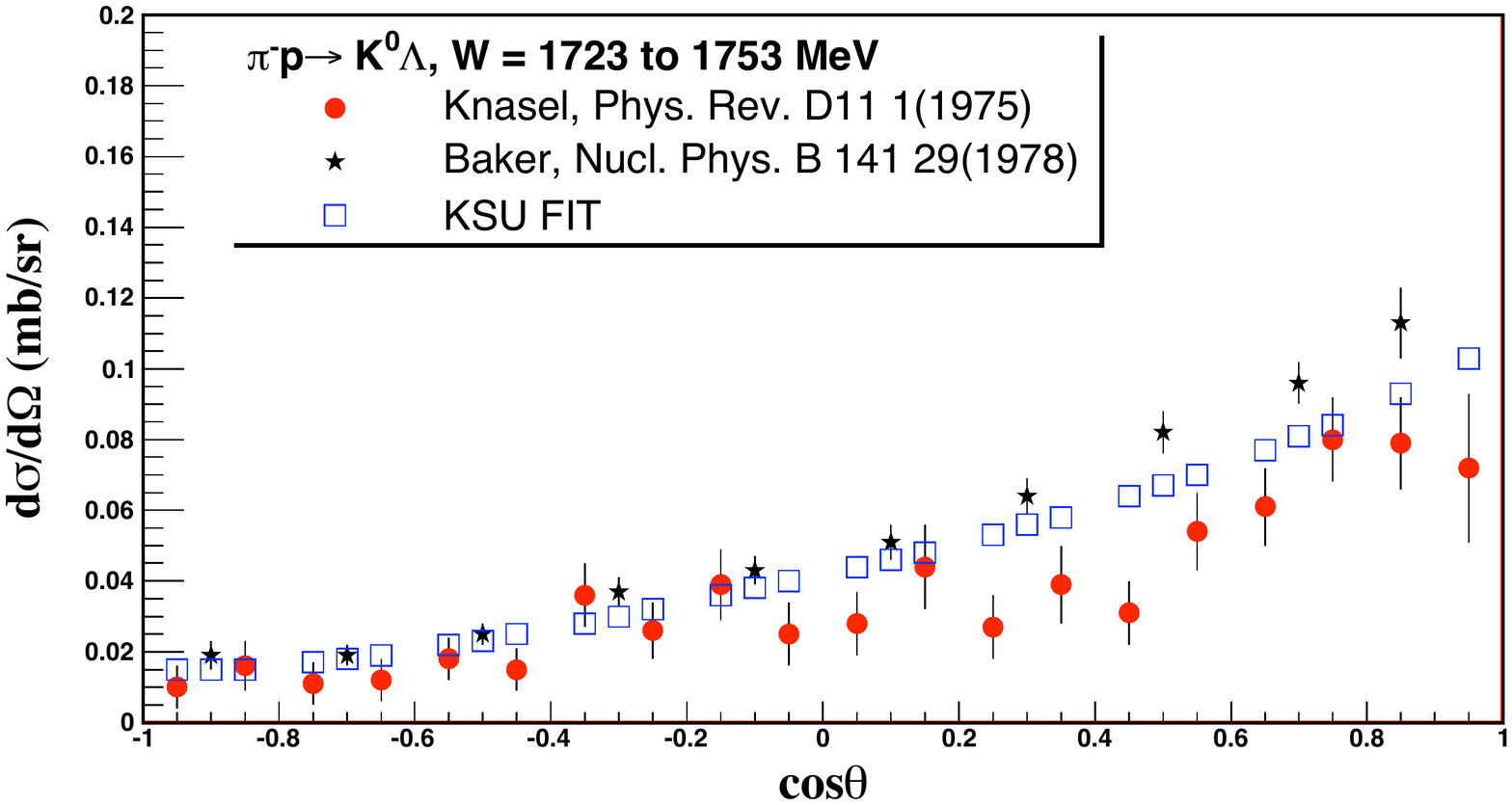}}
\vspace{-49mm}
\vspace{-15mm}
\scalebox{0.6}{\includegraphics{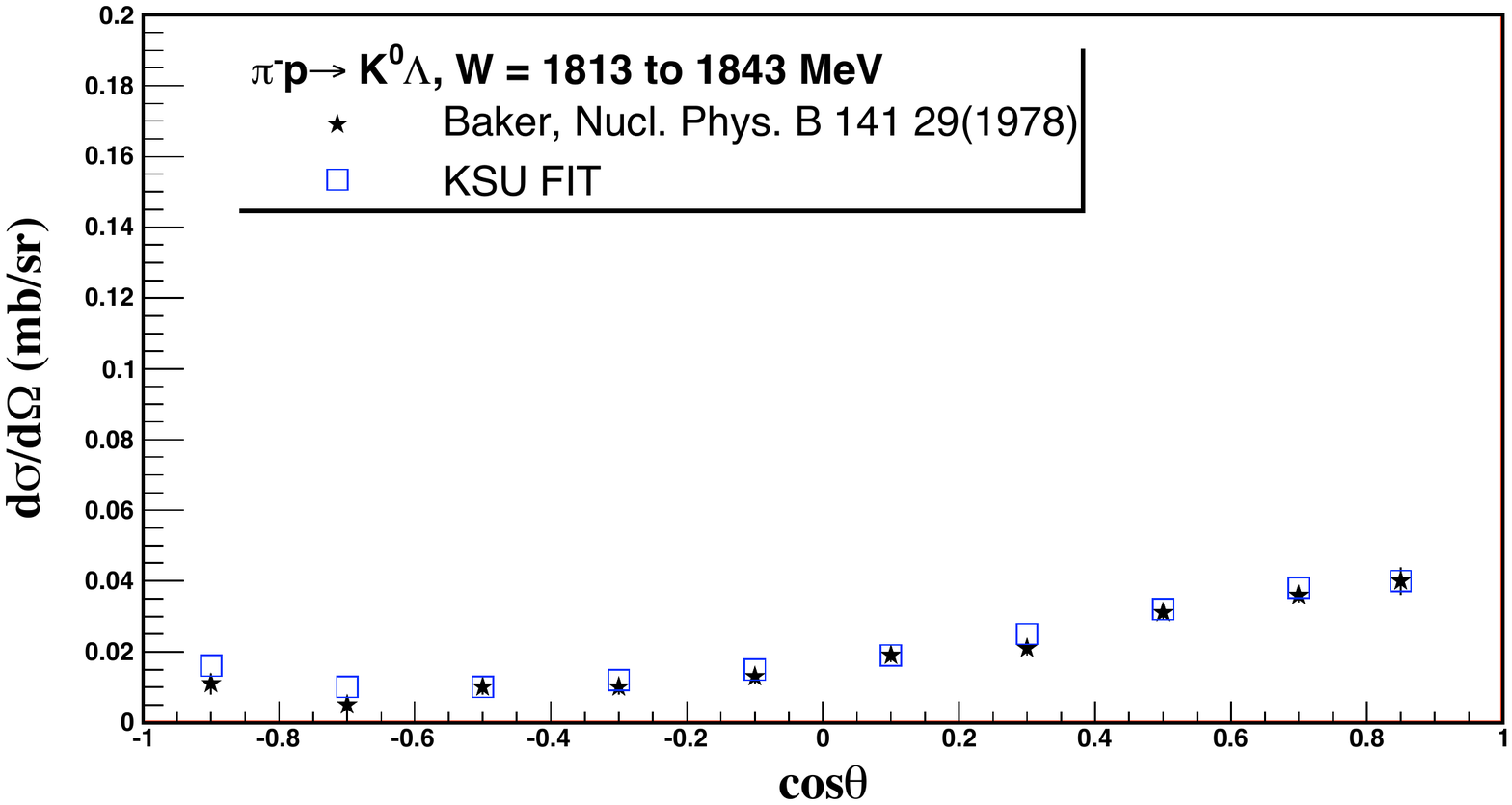}}
\vspace{-15mm}
\caption{Single-energy fit results for $\pi^- p\rightarrow K^0 \Lambda$.}
\label{fig:Amplitudes}
\end{figure*}

\begin{figure*}
\vspace{-45mm}
\addtocounter{figure}{-1}
\scalebox{0.6}{\includegraphics{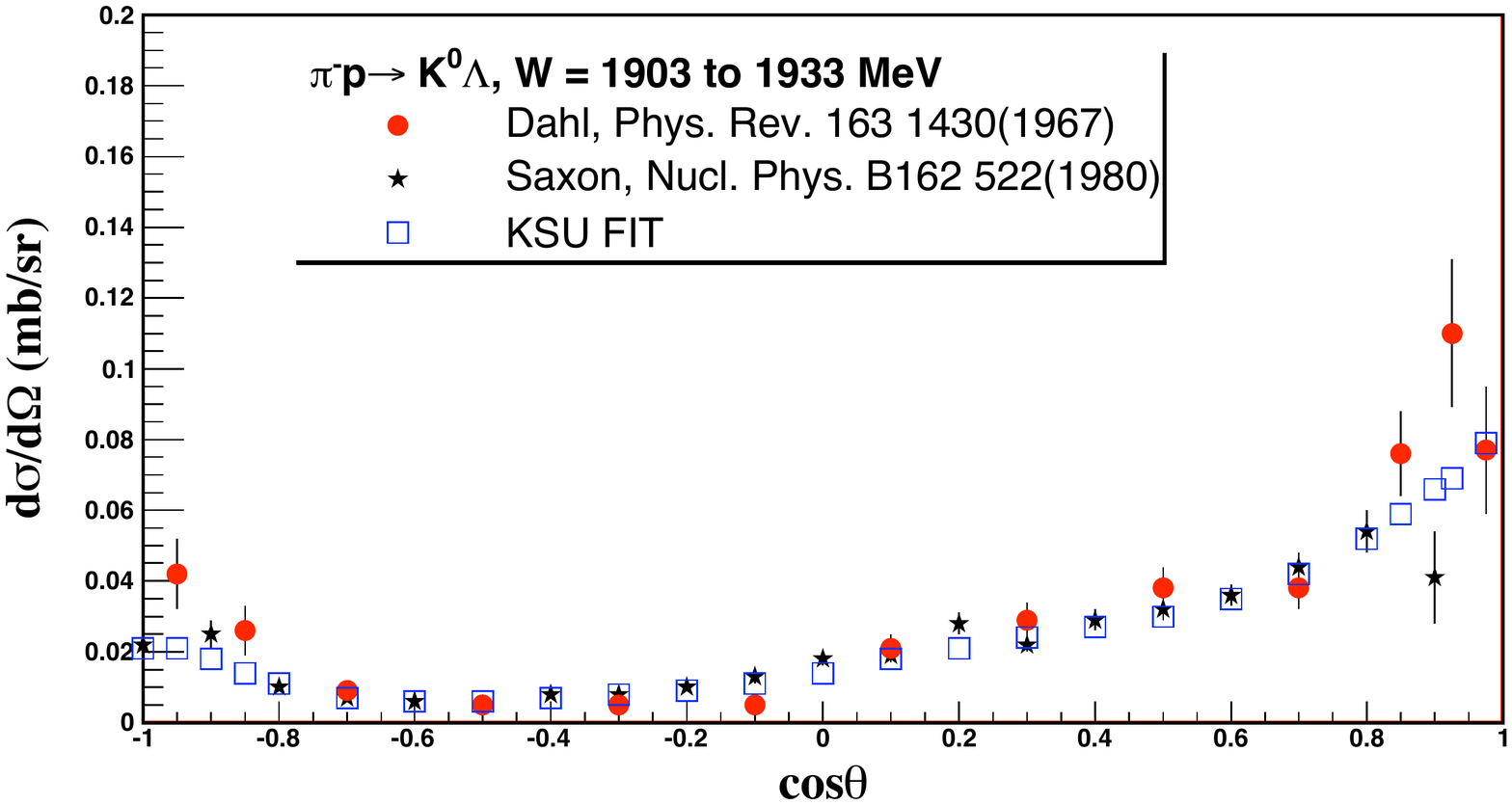}}
\vspace{-31mm}
\vspace{-33mm}
\scalebox{0.6}{\includegraphics{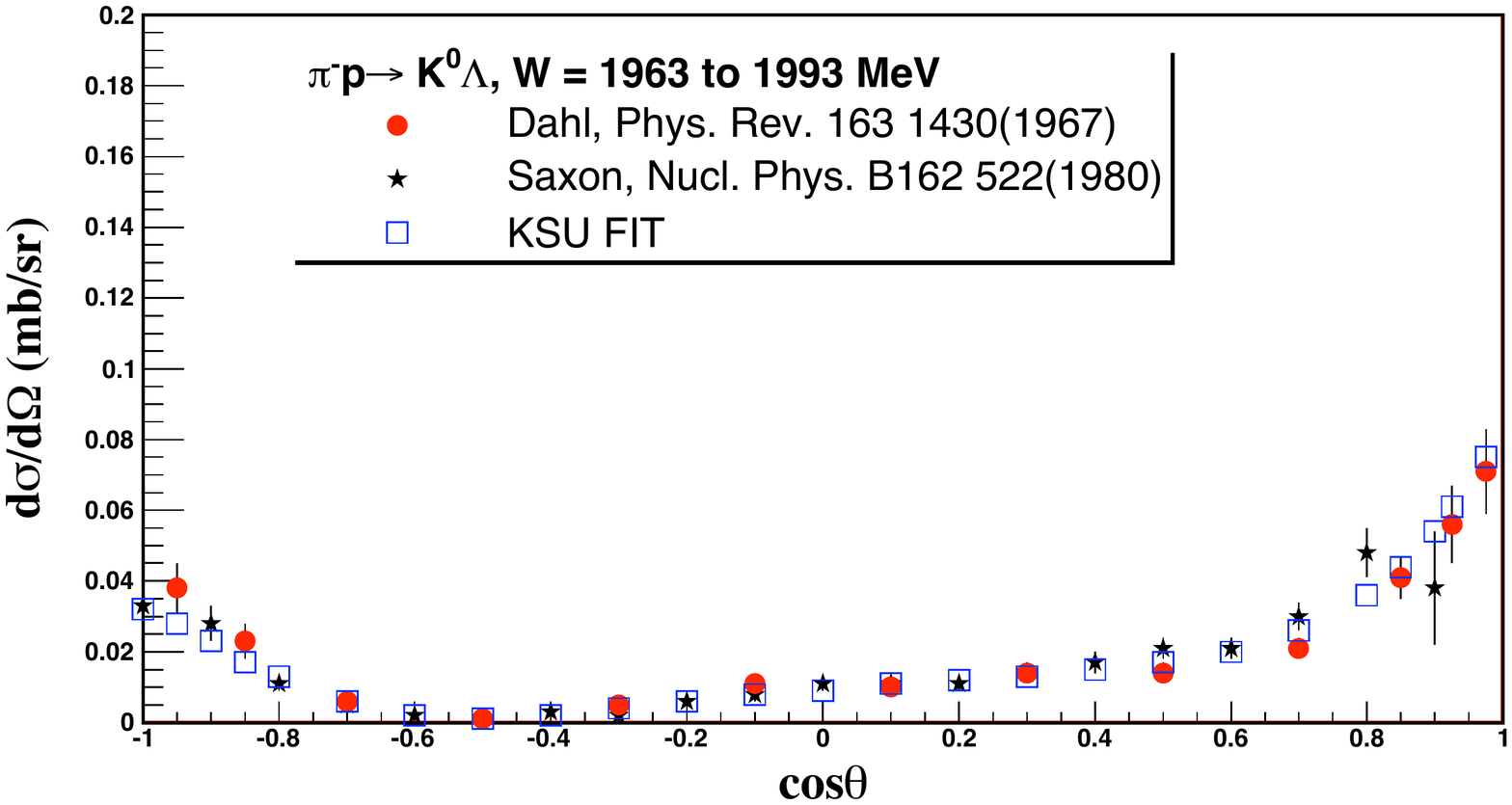}}
\vspace{-30mm}
\vspace{-33mm}
\scalebox{0.6}{\includegraphics{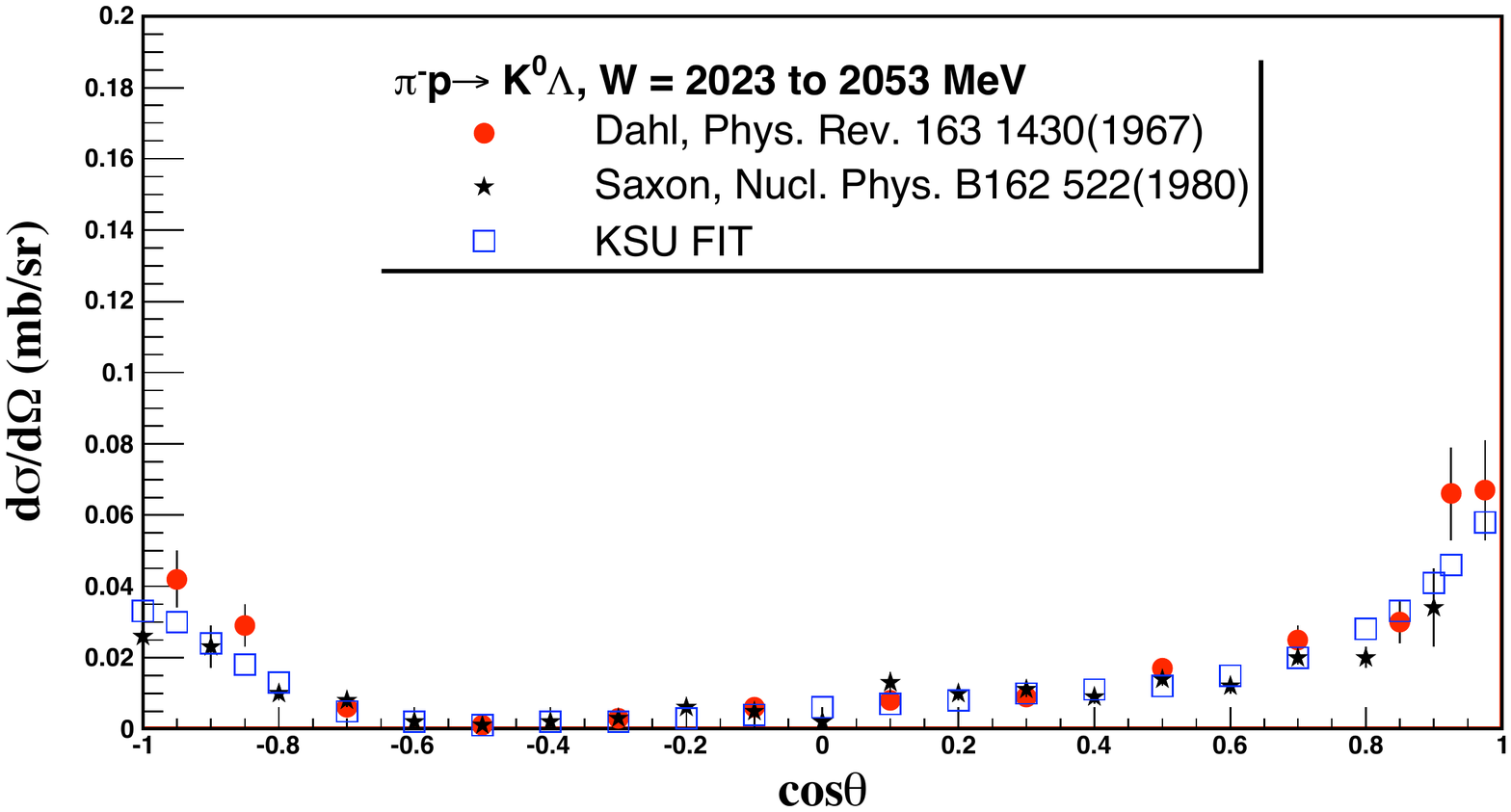}}
\vspace{-15mm}
\caption{(Cont'd)}
\label{fig:Amplitudes}
\end{figure*}

\begin{figure*}[htpb]
\vspace{-45mm}
\scalebox{0.45}{\includegraphics{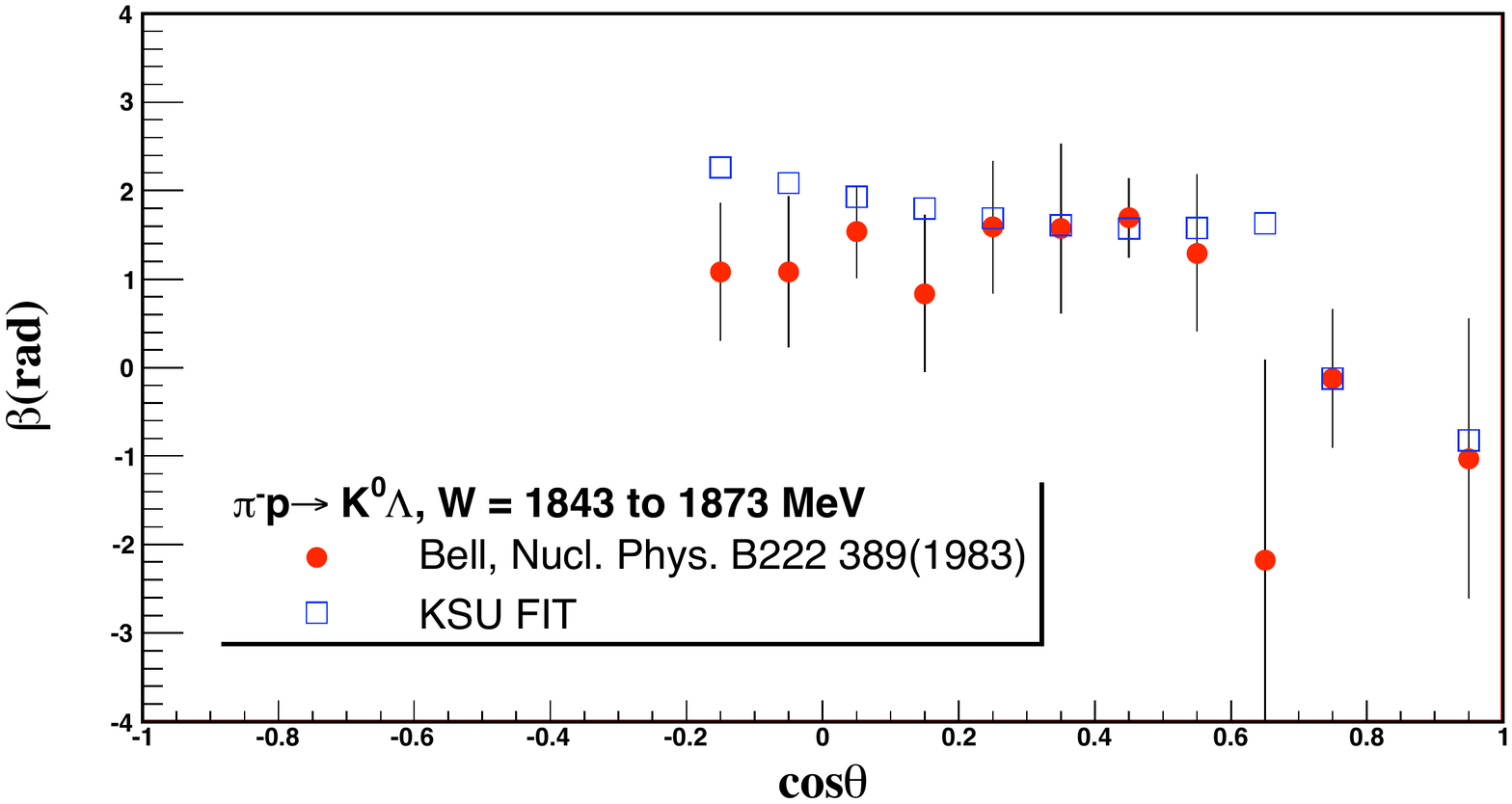}}
\vspace{-5mm}
\vspace{-35mm}
\scalebox{0.45}{\includegraphics{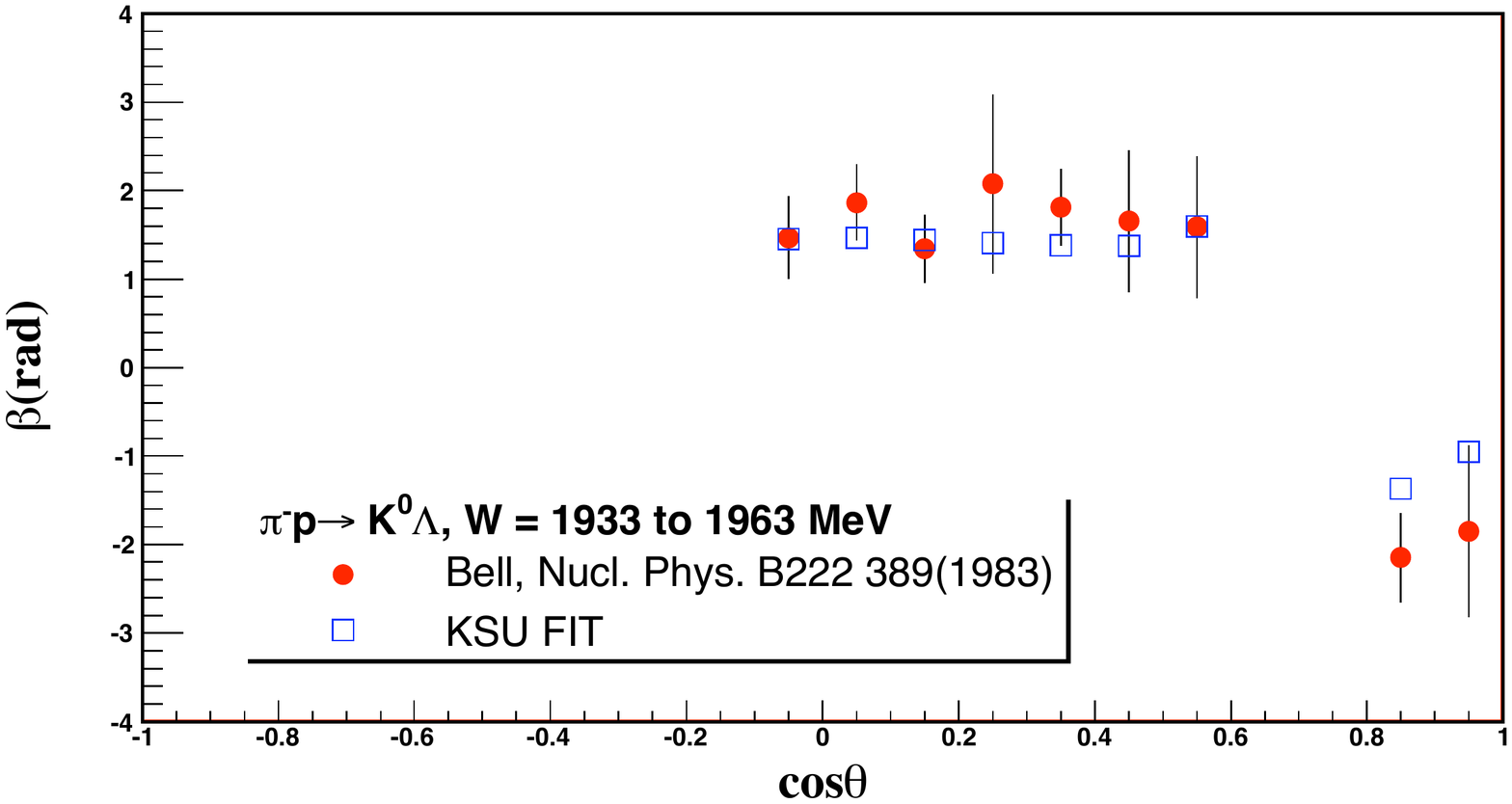}}
\vspace{-25mm}
\vspace{-15mm}
\scalebox{0.45}{\includegraphics{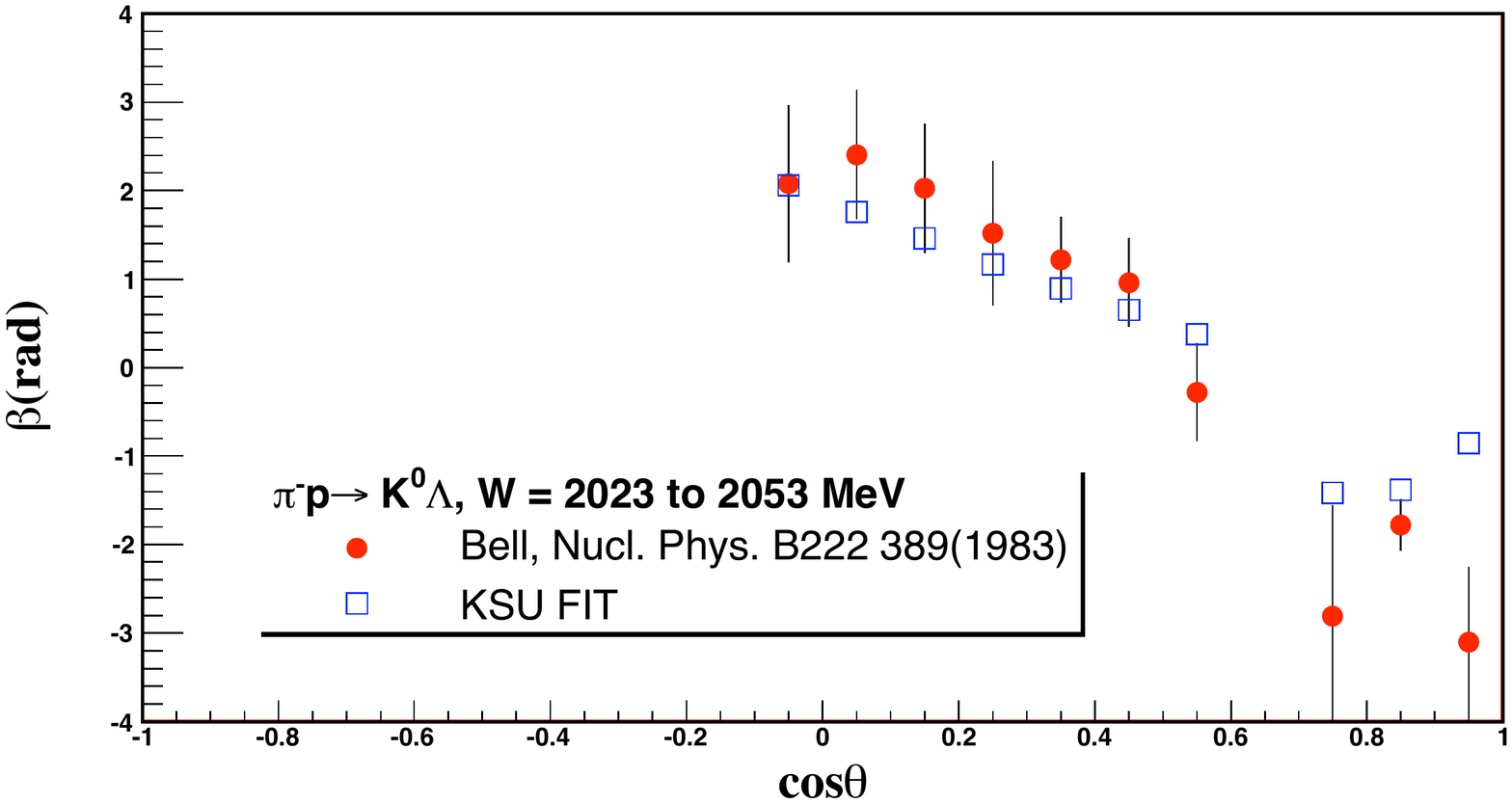}} 
\vspace{-25mm}
\vspace{-15mm}
\scalebox{0.45}{\includegraphics{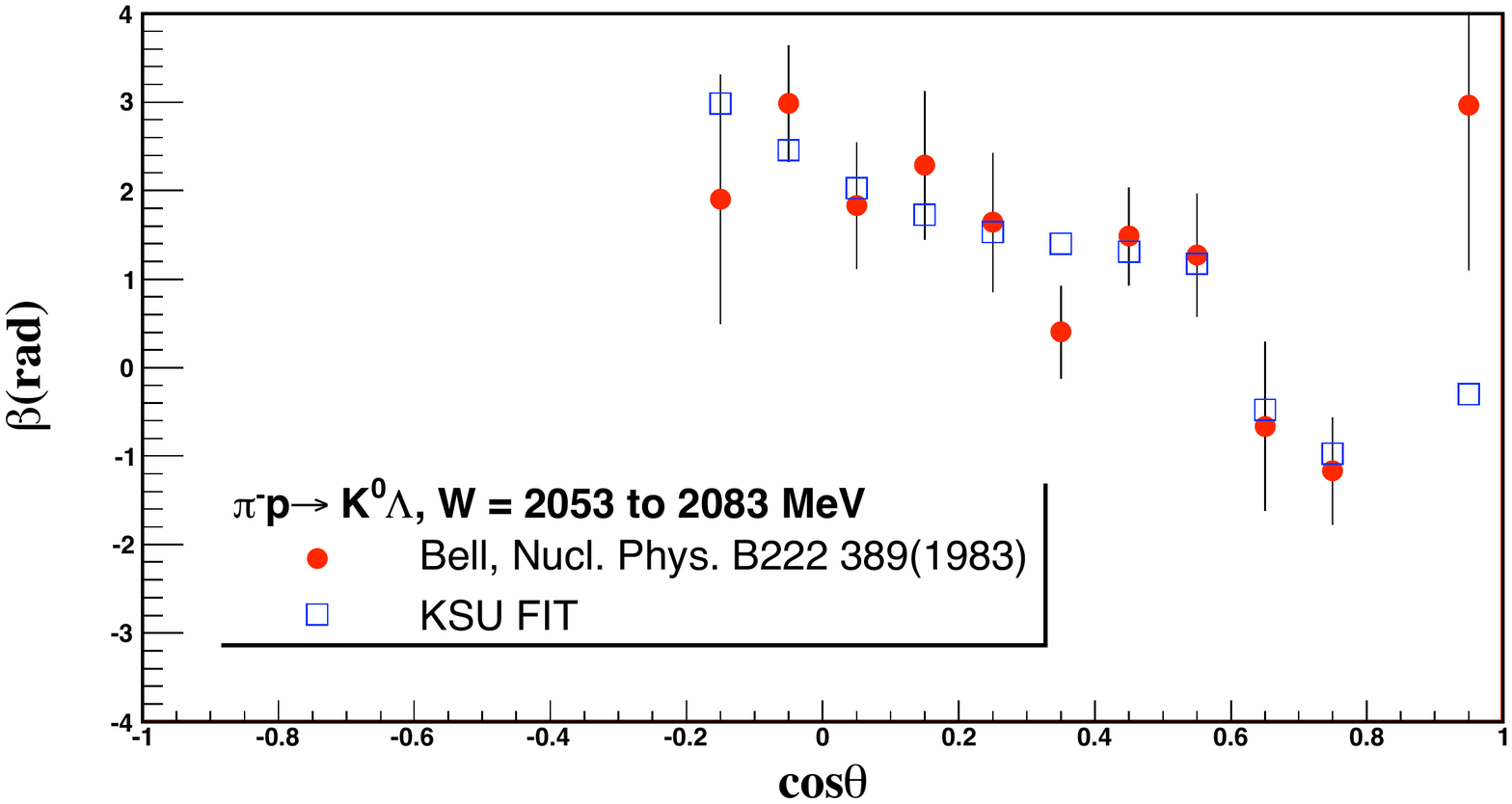}}
\caption{Single-energy fit results for $\pi^- p\rightarrow K^0 \Lambda$.}
\label{fig:Amplitudes}
\end{figure*}


\begin{figure*}[htpb]
\vspace{-15mm}
\vspace{-5mm}
\scalebox{0.6}{\includegraphics{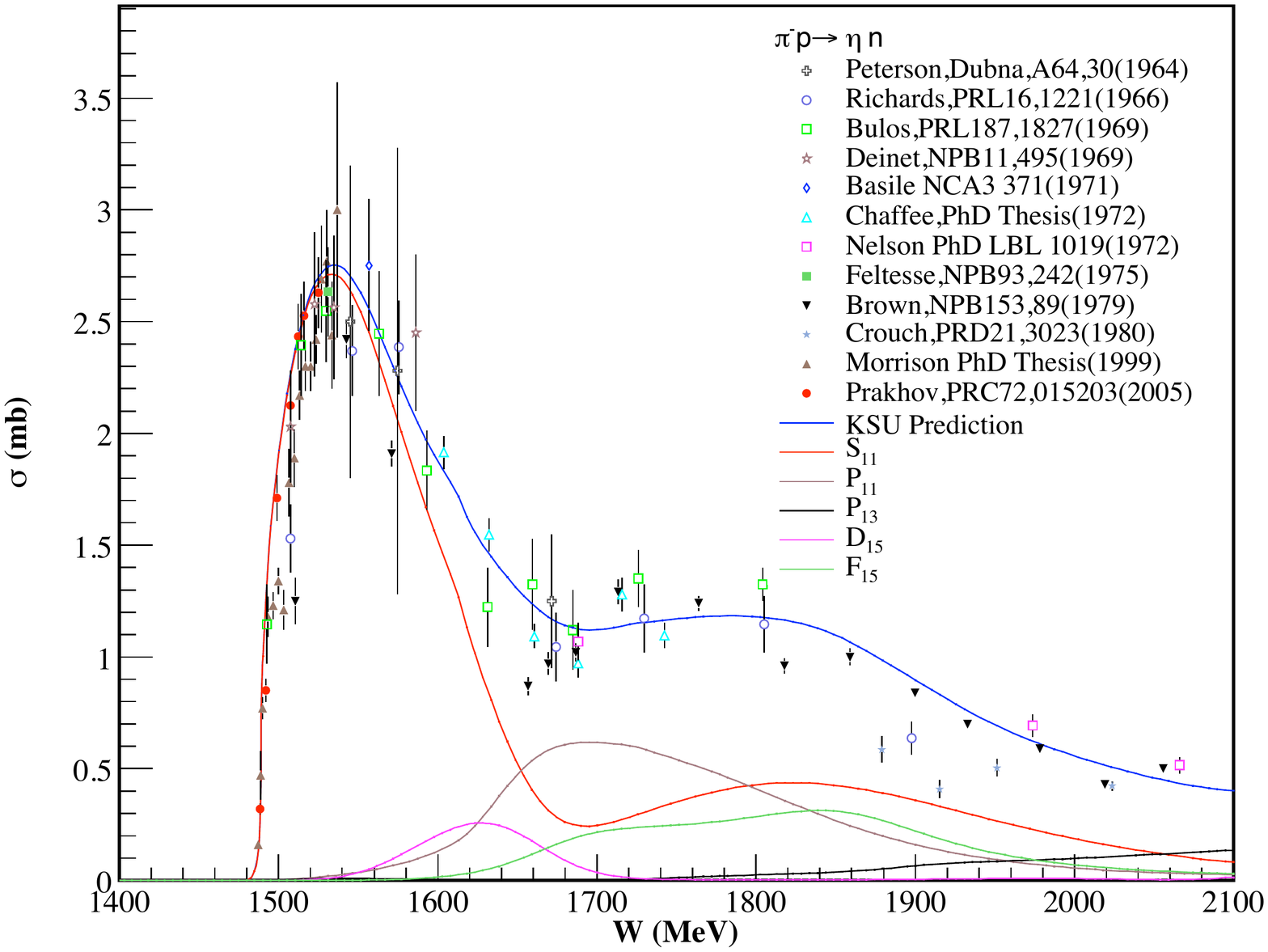}}
\vspace{-10mm}
\scalebox{0.6}{\includegraphics{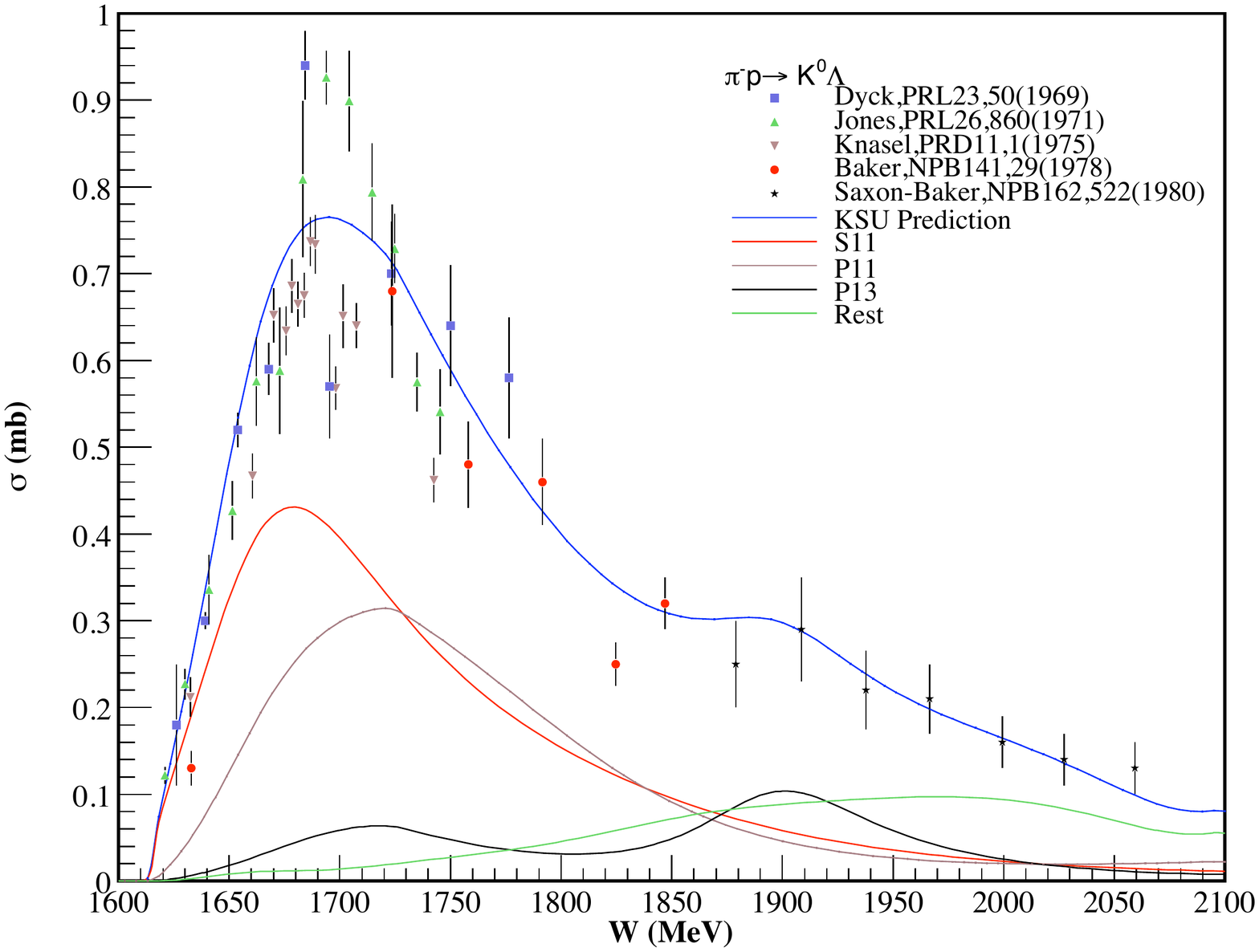}}
\vspace{-15mm}
\caption{ Predictions of integrated cross sections of $\pi^- p\rightarrow \eta n$ and $\pi^- p\rightarrow K^0 \Lambda$. }
\label{fig:Amplitudes}
\end{figure*}

\section{\emph{\bf  Summary and Conclusions}}
We have extracted partial-wave amplitudes for $\pi N\rightarrow \eta N$ and $\pi N\rightarrow K\Lambda$ from a constrained single-energy analysis from threshold to a c.m.\ energy of 2.1 GeV. The contributing partial waves for $\pi N\rightarrow \eta N$ were found to be $S_{11}$, $P_{11}$, $P_{13}$, $D_{15}$, $F_{15}$, and $G_{17}$. For $\pi N\rightarrow K\Lambda$, $S$-,  $P$-, and $D$-waves alone were sufficient to describe the differential cross section and polarization data  but additional small partial waves ($F_{15}$, $F_{17}$, $G_{17}$, $G_{19}$, and $H_{19}$) were necessary to obtain a good fit of the spin-rotation data. 

 In conclusion, we have investigated $\pi N\rightarrow \eta N$ and $\pi N\rightarrow K\Lambda$ reactions through single-energy analyses constrained by a global unitary energy-dependent fit. Our results for $\pi N\rightarrow K\Lambda$ are mostly consistent with the analysis by Bell {\it et al.\ }\cite{bell83} and with the Bonn-Gatchina analysis \cite{sarantsev11}. The inclusion of these amplitudes, in addition to $\pi N$, $\pi\pi N$, and $\gamma N$ into the global fit yields highly constrained information on resonance couplings. Also predictions of the integrated cross sections for $\pi^-p\rightarrow \eta n$ and $\pi^-p\rightarrow K^0\Lambda$ from the final global energy-dependent solution are in excellent agreement with the available data. 


\acknowledgements{This work was supported by the U.S. Department of Energy Grant No. DE-FG02-01ER41194. The authors thank the GWU group and especially Igor Strakovsky for providing part of the database for $\pi^-p\rightarrow \eta n$}.



\begin{thebibliography}{99}
  \bibitem{feltesse75} J. Feltesse {\textit{et al.}},  Nucl. Phys. B {\textbf{93}}, 242 (1975).
 \bibitem{baker79} R.D. Baker {\textit{et al.}},  Nucl. Phys. B {\textbf{156}}, 93 (1979).
 \bibitem{knasel75} T.M. Knasel {\textit{et al.}}, Phys. Rev. D {\textbf{11}}, 1 (1975).
 \bibitem{baker78} R.D. Baker {\textit{et al.}},  Nucl. Phys. B {\textbf{141}}, 29 (1978).
 \bibitem{saxon80} D.H. Saxon {\textit{et al.}},  Nucl. Phys. B {\textbf{162}}, 522 (1980).
 \bibitem{bell83} K.W. Bell {\textit{et al.}},  Nucl. Phys. B {\textbf{222}}, 389 (1983).
\bibitem{manley92} D.M. Manley and E.M. Saleski, Phys. Rev. D {\textbf{45}}, 4002 (1992).
\bibitem{arndt06} R.A. Arndt, W.J. Briscoe, I.I. Strakovsky, and R.L. Workman, Phys. Rev. C {\textbf{74}}, 045205 (2006).
\bibitem{manoj12} M. Shrestha, and D.M. Manley, to be submitted to Phys. Rev. C. 
\bibitem{prakhov05} S. Prakhov {\textit{et al.}}, Phys. Rev. C {\textbf{72}}, 015203 (2005).
 \bibitem{debeham75} N.C. Debeham {\textit{et al.}}, Phys. Rev. D {\textbf{12}}, 2545 (1975).
 \bibitem{richards70} W.B. Richards {\textit{et al.}}, Phys. Rev. D {\textbf{1}}, 10 (1970).
 \bibitem{deinet69} W. Deinet {\textit{et al.}},  Nucl. Phys. B {\textbf{11}}, 495 (1969).
 \bibitem{brown79} R.M. Brown {\textit{et al.}},  Nucl. Phys. B {\textbf{153}}, 89 (1979).
 \bibitem{dahl67} Orin I. Dahl, {\textit{et al.}}, Phys. Rev.  {\textbf{163}}, 1430 (1967).
 \bibitem{batinic95} M. Batini\' c {\textit{et al.}}, Phys. Rev. C {\textbf{51}}, 2310 (1995).
 \bibitem{sarantsev11} A. Sarantsev, Sixth International Workshop on $\pi N$ Partial-Wave Analysis, GWU, May 23 - 27, 2011- http://gwdac.phys.gwu.edu/pwa2011/Tuesday/sarantsev\_PWA2011.pdf
 \bibitem{aniso11} A.V. Anisovich {\textit{et al.}}, Eur. Phys. Jour. A {\textbf{47}}, 153 (2011).
 \bibitem{durand08} J. Durand {\textit{et al.}}, Phys. Rev. C {\textbf{78}}, 025204 (2008).
 \bibitem{julich12} K. Nakayama {\textit{et al.}}, J. Korean Phys. Soc. {\textbf{59}}, 224 - 246 (2011). 
 
\end{thebibliography}
  \end{document}